# Correlations between the structural, magnetic, and ferroelectric properties of BaMO$_3$: M = Ti$_{1-x}$(Mn/Fe)$_x$ compounds: A Raman study

Bommareddy Poojitha, Ankit Kumar, Anjali Rathore, and Surajit Saha*

Department of Physics, Indian Institute of Science Education and Research Bhopal, 462066, India

**Abstract:**

Multiferroics possess two or more switchable states such as polarization, magnetization, etc. Phonon excitations in multiferroic phase are strongly modified by magnetoelectric coupling, spin-phonon coupling, and anharmonic phonon-phonon interactions. Here, we have investigated the correlation between phonons and multiferroic order parameters in hexagonal BaMO$_3$: M = Ti$_{1-x}$(Mn/Fe)$_x$ systems using powder x-ray diffraction (PXRD), Raman spectroscopic, and magnetic measurements. The structural transformation from a polar tetragonal to a non-polar 6H-type hexagonal phase is observed as a function of doping (Mn/Fe). Magnetic measurements reveal that the BaTi$_{1-x}$Mn$_x$O$_3$ is paramagnetic while BaTi$_{1-x}$Fe$_x$O$_3$ exhibits composition-dependent ferromagnetic order. Importantly, Anomalous temperature-dependence is observed for two phonons ($E_{1g}$ at ~ 152 cm$^{-1}$ and $A_{1g}$ at ~ 636 cm$^{-1}$) in both the systems exhibiting similar trend with the doping (Mn/Fe) irrespective of the differences in their magnetic ground state. Hence, we attribute the phonon anomalies in both the (Fe/Mn doped) systems to strong anharmonic phonon-phonon interactions arising from large atomic displacements involved in the vibrations. In addition, we have also observed signatures of correlation of phonons with ferroelectric phase as well as magnetically ordered state suggesting the presence of a strain-induced magnetoelectric coupling in the doped compounds.

**Keywords:** Composite materials, Structure, Ferroelectrics, Magnetically ordered materials, Solid-state reactions, Phonons.

**Introduction:**

Magnetoelectric multiferroics, exhibiting coupled ferroelectric and magnetic order parameters, show great potential for emerging technological applications and are also an important platform for the study of the rich physics governing their physical properties [1-3]. The search for multiferroic materials has drawn a renewed interest in recent years after the predictions of strong magnetoelectric effect in ferromagnet-ferroelectric composite systems:

Fe/BaTiO$_3$ [4], Fe$_3$O$_4$/BaTiO$_3$ [5], and Co$_2$MnSi/BaTiO$_3$ [6,7]. The preparation of composite materials with strongly coupled multiferroic properties near room temperature could be a turning point for modern electronics and multifunctional materials. For instance, new multiferroic materials can be fabricated by introducing magnetic ions into conventional ferroelectrics [8]. BaTiO$_3$ is a polymorphic system exhibiting two types of crystal structures, perovskite (tetragonal) and 6H-type hexagonal at room temperature, where the perovskite phase is more stable near room temperature showing ferroelectricity. Importantly, the perovskite BaTiO$_3$ undergoes a series of structural transitions [9] as a function of temperature (T) from rhombohedral (at T <193 K) ➔ orthorhombic (at T <280 K) ➔ tetragonal (T<400 K) ➔ cubic (at T <1733 K) ➔ hexagonal (6H-BaTiO$_3$). BaTiO$_3$ is paraelectric in cubic phase and becomes ferroelectric in the low-temperature phases having electrical polarization along [001], [011], and [111] axes (denoted in terms of cubic notation) of the tetragonal, orthorhombic, and rhombohedral unit cell, respectively [9]. Room temperature 6H-phase of BaTiO$_3$ in single crystal form were synthesized by optical floating zone technique exhibiting a ferroelectric phase below 74 K [10]. Stabilizing the chemically undoped 6H-phase of BaTiO$_3$ includes several complications [11,12]. Surprisingly, the 6H phase can be stabilized at room temperature even under ambient conditions by simply doping 3$d$ transition metal ions at the Ti site [13-17]. The novel material BaTiO$_3$ upon appropriate doping exhibits conspicuous properties such as giant electro-strain effect, photorefractive effect, and magneto-optical properties for waveguide-based optical isolators [18-21]. They are currently being addressed to investigate as a family of contemporary ABO$_3$ multiferroic materials [22,23]. Inducing magnetism in classical ferroelectrics such as BaTiO$_3$ has been of great importance from the fundamental point of view as well as for realizing applications. It was theoretically predicted by Hiroyuki Nakayama et al. that Cr, Mn, and Fe are good candidates to stabilize ferromagnetism in BaTiO$_3$ system [8]. Further, introduction of charge along with Mn doping may induce ferromagnetism (FM) which otherwise gives antiferromagnetic (AFM) ordering [8]. On the other hand, Xu et al. [17] have predicted multiferroicity in Fe doped BaTiO$_3$.

Similar to BaTiO$_3$, both BaMnO$_3$ [24-26] and BaFeO$_3$ [27,28] compounds are also polymorphic, exhibiting even richer phase diagrams with a variety of crystal structures which are very sensitive to and tunable by synthesis conditions and/or chemical doping. For example, BaMnO$_3$ can be stabilized in various phases like 2H, 15R, 9R, 12R, 6H, 4H, etc. [24-26] by tuning oxygen stoichiometry and similarly, BaFeO$_3$ can exist in monoclinic, 6H, triclinic, rhombohedral, tetragonal and cubic phases [27,28]. For instance, 6H-BaMnO$_{3-\delta}$ and 6H-BaFeO$_{3-\delta}$ are antiferromagnetic with T$_N$ ~ 250 K [26] and 130 K [29], respectively. In 6H-BaMnO$_{3-\delta}$, spins are aligned antiparallel to each other along the *c*-axis below T$_N$ (ferromagnetically ordered in a single layer with no simple superexchange pathways between the cations in the layer) [26]. However, the magnetic ground state of 6H-BaFeO$_{3-\delta}$ is very complex including FM, AFM, and paramagnetic interactions over a wide range of temperatures [29]. Further, the 6H-type BaFeO$_{3-\delta}$ was reported to show a strong magnetodielectric (MD) effect which arises due to negative magnetoresistance and the Maxwell-Wagner interfacial effects [30]. Recently, Rubavathi et al. observed MD effect in Fe substituted BaTiO$_3$ even at lower concentrations (5% and 10% Fe) [31]. It was reported that the partial substitution of Ti$^{+4}$ in BaFeO$_{3-\delta}$ increases Fe$^{+4}$ concentration and oxygen stoichiometry and hence, enhances the FM ordering [30]. These novel phenomena in BaMO$_3$ have led to several studies on transition metal doped BaTiO$_3$ materials but most of these are dedicated to explore structural phase transitions and magnetic properties [17,30,32-38]. To the best of our knowledge, detailed study on the phonons of these systems and their roles in the physical properties is still lacking. We have previously shown that the non-magnetic Eu$^{+2}$ ions at Ba site of BaTiO$_3$ lattice suppresses the anharmonic phonon-phonon interactions, and hence, ferroelectricity [39]. Since phonons play a vital role in understanding classical ferroelectric materials, one may expect them as key-player in understanding the magnetoelectric multiferroics too.

Motivated by the possibilities of phonon-engineering of these systems, we have attempted to induce magnetism in BaTiO$_3$ lattice by doping transition metals such as Mn/Fe at Ti sites and study the possible correlation between phonons and ferroelectric (multiferroic) order parameter(s) through x-ray diffraction, magnetic, and Raman spectroscopic measurements. Both the systems undergo a structural transformation from tetragonal to 6H phase upon doping (Mn/Fe). Two characteristic phonons of the 6H phase, i.e. the $E_{1g}$ mode at ~ 152 cm$^{-1}$ and $A_{1g}$ mode at ~ 636 cm$^{-1}$ are found to exhibit anomalous behaviour with temperature in the low range of doping by Mn/Fe in BaTiO$_3$ i.e. in BaTi$_{0.98}$Mn$_{0.02}$O$_3$ and BaTi$_{0.95}$Fe$_{0.05}$O$_3$. Magnetic studies reveal that the BaTi$_{1-x}$Mn$_x$O$_3$ is paramagnetic while BaTi$_{1-x}$Fe$_x$O$_3$ exhibits composition-sensitive ferromagnetic ordering. Surprisingly, the phonon anomaly reduces with further Mn/Fe-doping in both the systems irrespective of their magnetic ground states. Hence, the anomalous temperature-dependence of phonons is attributed to strong phonon-phonon anharmonic interactions. The signatures of correlation between phonons and ferroelectricity, as well as magnetoelectric coupling are discussed based on temperature-dependent Raman, magnetization, and x-ray diffraction measurements.

**Experimental Details**

Polycrystalline samples of $BaTi_{1-x}Mn_xO_3$ (x = 0.02, 0.05, 0.10, 0.15, 0.20, 0.30), and $BaTi_{1-x}Fe_xO_3$ (x = 0.05, 0.10, 0.125, 0.15, 0.20, 0.25) were synthesized by using high temperature solid-state reaction method. High purity powder chemicals of $BaCO_3$, $TiO_2$, $MnO_2$, and $Fe_2O_3$ were used as precursors. The stoichiometric mixed powder was calcined at 1250°C and 1300°C and sintered at 1350°C for 2 hours each with intermediate grindings. Powder x-ray diffraction (PXRD) patterns were collected by using PANalytical Empyrean x-ray diffractometer with Cu-$K_\alpha$ radiation of wavelength 1.5418 Å. Chemical compositions were determined using energy dispersive x-ray (EDAX) technique equipped with high-resolution field emission scanning electron microscope (HR-FESEM) (Zeiss ULTRA Plus). Raman spectra were recorded on pellets of $BaTi_{1-x}Mn_xO_3$, and $BaTi_{1-x}Fe_xO_3$ using a LabRAM HR Evolution Raman spectrometer equipped with Peltier cooled charge-coupled detector (CCD) and Nd:YAG laser of 532 nm wavelength as the source of excitation. The dc magnetization measurements below room temperature were carried out in Quantum-Design Superconducting Quantum Interference Device attached with Vibrating Sample Magnetometer (SQUID-VSM). A Quantum-Design PPMS (Physical Property Measurement System) attached with VSM, oven, and ultra-high vacuum facilities was used for dc magnetization measurements in the high-temperature range (300K – 800 K).

**Results and discussion**

Figure 1 shows the crystal structure of tetragonal and 6H-type hexagonal unit cells of $BaMO_3$: M = $Ti_{1-x}(Mn/Fe)_x$ drawn using VESTA software [40]. The atomic arrangement in tetragonal unit cell is similar to that in the ideal cubic structure where the $TiO_6$ octahedra are connected to each other through octahedral corners. Doping of Mn/Fe at the Ti site increases the tolerance factor $\left[t = \frac{r_A+r_O}{\sqrt{2}(r_B+r_O)}\right]$ [41] and this leads to tetragonal to hexagonal structural phase transition. The 6H structure consists of dimers of $M_2O_9$ units (two face-shared octahedra) separated by a single layer of cubic perovskite-type octahedra that corner-share with the $M_2O_9$ units. The stacking sequence of the atoms in 6H-phase is $(cch)_2$ where c and h represent cubic (corner-shared octahedra) and hexagonal (face-shared octahedra) layers, respectively. Each of the Ba, M, and O atoms has two inequivalent crystallographic sites which are labelled as Ba1, Ba2, M1, M2, O1, and O2, respectively [42] (Figure 1). The metal-metal (M-M) bond along the *c*-axis is remarkable in 6H phase as the M-M separation is significantly short, ~ 2.72 Å (which is comparable to 2.46 (2.47) Å in metallic Fe (Mn)), which may lead to off-centred displacement of M2 cations due to the strong electrostatic repulsion between them (Figure 1). As a result, all the six M1-O bonds are identical in corner-shared M1O6 octahedra, while the face-shared M2O6 octahedra are strongly distorted with three long and three short M2-O bond lengths [42].

Powder x-ray diffraction (PXRD) pattern for $BaTi_{1-x}Mn_xO_3$ and $BaTi_{1-x}Fe_xO_3$ compounds collected at room temperature are shown in Figure 2(a,b). All PXRD patterns were refined with Rietveld method using High Score Plus software. The refinement results reveal that $BaTi_{0.98}Mn_{0.02}O_3$ and $BaTi_{1-x}Fe_xO_3$: x = 0.05 to 0.15 are stabilized in mixed-phase composed of tetragonal structure with space group P4mm (No.99) and 6H structure with space group $P6_3mmc$ (No. 194). The weight fractions of both the phases and lattice parameters at each composition are extracted from the double-phase refinement method using High Score Plus software (see Figure S1 and Table S1 in section S1 of supplementary material). The phase-fraction of tetragonal structure gradually decreases and that of hexagonal (6H) structure increases from x: 0.05 to 0.15 in $BaTi_{1-x}Fe_xO_3$ which implies that the structural transition from tetragonal to 6H type hexagonal phase as a function of doping in these systems is of second order in nature. On the other hand, the higher compositions i.e $BaTi_{1-x}Mn_xO_3$: x = 0.05 to 0.30, and $BaTi_{1-x}Fe_xO_3$: x = 0.20 to 0.25 are stabilized in pure single-phase 6H structure [43]. The lattice parameters corresponding to 6H unit cell ($a_{6H}$ and $c_{6H}$) in $BaTi_{1-x}Mn_xO_3$ are observed to decrease with Mn concentration as shown in Figure 2(c). This can be attributed to smaller ionic radius [44] of $Mn^{+4}$ ($r_{Mn+4}$ = 0.530 Å) ions than $Ti^{+4}$ ($r_{Ti+4}$ = 0.605 Å) ions thus also implying that the concentration of $Mn^{+4}$ is significantly higher than that of $Mn^{+3}$ ions ($r_{Mn+3}$ = 0.645 Å with high-spin configuration), if present. On the other hand, in the $BaTi_{1-x}Fe_xO_3$ system, the in-plane lattice parameter ($a_{6H}$) exhibits an increasing trend with Fe content till x = 0.15 and then decreases with further increment of the doping. However, the out-of-plane parameter ($c_{6H}$) increases with increasing Fe content throughout the investigated composition range (x: 0.05 to 0.25) (Figure 2(d)). It should be noted that the $Fe^{+4}$ (0.585 Å) is also smaller than $Ti^{+4}$ (0.605 Å) which should lead the lattice to shrink and, hence, decrease the lattice parameters. In contrast, the otherwise observed behaviour implies that the concentration of $Fe^{+3}$ ions in our $BaTi_{1-x}Fe_xO_3$ compounds is significant which causes the unit cell to expand due to its larger ionic radius ($r_{Fe+3}$ = 0.645 Å) than that of $Ti^{+4}$. The slope reversal in the $a_{6H}$ vs $x$ curve around x ~ 0.15 can be addressed on the basis of crystal structure and strain-induced ferromagnetism. It should be noted here that the compounds have two coexisting crystallographic phases in the composition range of x = 0.05 to 0.15 for $BaTi_{1-x}Fe_xO_3$ that results in a strain imposed on the unit cells of both the phases across the phase boundaries which is absent for x > 0.15 due to the presence of single-phase 6H structure at higher Fe-doping. Interestingly, the coercive field in the magnetic data, which will be discussed in detail later, also shows a similar trend as that of the $a_{6H}$ vs $x$. The lattice parameters corresponding to the tetragonal phase ($a_T$ and $c_T$) are shown in Figure S1. It is to be noted that the in-plane (out of plane) parameter decreases (increases) with increasing Fe concentration. This indicates that the dopants get incorporated in both the tetragonal and 6H unit cells. Further, chemical compositions were determined using energy dispersive x-ray (EDAX) and found to be matching well with the expected stoichiometry for all compositions of the both systems ($BaTi_{1-x}Mn_xO_3$ and $BaTi_{1-x}Fe_xO_3$) within the experimental resolution (see Figure S2 in Section S2 of supplementary material for complete details).

Raman spectra of $BaTi_{1-x}Mn_xO_3$ and $BaTi_{1-x}Fe_xO_3$ for all the compositions are collected at room temperature and shown in Figure 3 (a,d) (full spectral range is shown in Figure S3 of supplementary material). The tetragonal $BaTiO_3$ has five atoms per unit cell giving rise to 12 optical phonons and eight of them are Raman active with irreducible representations $\Gamma_{Raman} = 3A_1+B_1+4E$ [45]. Since Raman spectroscopy is a local probe and our focus is mainly on the phonons corresponding to 6H phase in this article, we have collected Raman spectra in such a way that the spots corresponding to the 6H phase is probed without getting interfered by the presence of tetragonal phase. Still, one can see the presence of a mode at around ~ 308 cm$^{-1}$ in the mixed phase compounds which is a signature Raman mode of the tetragonal phase [46-51]. The factor group analysis predicts nineteen Raman active phonons for 6H-type hexagonal structure with space group $P6_3mmc$ that can be classified as $\Gamma_{Raman} = 5A_{1g}+6E_{1g}+8E_{2g}$ [45]. Phonon parameters are extracted by analysing the Raman spectra using Lorentzian functions. The well resolved modes for $BaTi_{1-x}Mn_xO_3$ with x = 0.05 sample at room temperature are 71 [$E_{2g}$], 104 [$A_{1g}$], 126 [$E_{2g}$], 152 [$E_{1g}$], 174 [$E_{2g}$], 203 [$E_{2g}$], 219 [$E_{1g}$], 395 [$E_{2g}$], 482 [$E_{2g}$], 635 [$A_{1g}$], 691, 714, and 802 [$A_{1g}$] cm$^{-1}$. Similarly, Raman active modes at 73 [$E_{2g}$], 107 [$A_{1g}$], 153 [$E_{1g}$], 223 [$E_{1g}$], 606 [$A_{1g}$], 637 [$A_{1g}$], 711 cm$^{-1}$ are clearly observed for $BaTi_{1-x}Fe_xO_3$ with x = 0.05. The $A_{1g}$ mode at 802 cm$^{-1}$ is present in $BaTi_{1-x}Mn_xO_3$ with x = 0.02 and 0.05 and vanishes at higher doping. Notably, this $A_{1g}$ mode is absent in $BaTi_{1-x}Fe_xO_3$ for all compositions. The observed additional modes are due to the incorporation of Mn/Fe into the BaTiO$_3$ lattice [52]. Low-frequency modes at 71 [$E_{2g}$], 104 [$A_{1g}$], and 152 [$E_{1g}$] cm$^{-1}$ are associated with the displacement of Ba atom involving Ba-O bonds. The 635 [$A_{1g}$] cm$^{-1}$ mode in BaTi$_{1-x}$Mn$_x$O$_3$ is related to O vibrations located in (Ti/Mn)O$_6$ octahedra, whereas the phonons associated solely with TiO$_6$ and FeO$_6$ octahedra in BaTi$_{1-x}$Fe$_x$O$_3$ are well resolved having different energies at 637 [$A_{1g}$], and 606 [$A_{1g}$] cm$^{-1}$, respectively. Figure 3 (b,c) shows the frequency as a function of doping (Mn/Fe) for a few selected phonons. In BaTi$_{1-x}$Mn$_x$O$_3$, the frequency of low energy modes increases with increasing Mn concentration that can be attributed to the observed lattice contraction arising due to smaller ionic radius of Mn$^{+4}$ (0.550 Å) than Ti$^{+4}$ (0.605 Å). However, the frequency of $A_{1g}$ (635 cm$^{-1}$) mode decreases with doping which can be understood as a consequence of heavier mass of Mn (54.94 amu) than Ti (47.68 amu) atoms [53]. Since $A_{1g}$ (635 cm$^{-1}$) mode is the convolution of two phonons associated with Ti and Mn atoms, it is expected that the change in atomic mass will dominate the change in phonon frequency with doping (Mn). Similarly, all the phonons in BaTi$_{1-x}$Fe$_x$O$_3$ are observed to undergo a redshift upon increasing Fe incorporation in place of Ti. Since Fe (55.85 amu) is heavier than Ti (47.68 amu) atom, the change in atomic mass leads to a decrease in the phonon frequency. At the same time, lattice expansion upon Fe doping, as evidenced from PXRD (inset of Figure 2d), also causes phonon frequency to decrease with doping owing to the presence of larger size Fe$^{+3}$ (0.645 Å) in place of Ti$^{+4}$ (0.605 Å). Further, Ba-O and M-O (M = Mn/Fe) bond lengths decrease (increase) with increasing Mn (Fe) concentration (Figure S4 and S5 in supplementary material).

To get a better insight of the phonons and their association with the properties of these compounds, we have performed temperature-dependent Raman spectroscopic measurements for both BaTi$_{1-x}$Mn$_x$O$_3$ (from 80-400 K) and BaTi$_{1-x}$Fe$_x$O$_3$ (from 80-800 K) systems. It is to be noted that the number of Raman modes remains the same in the investigated temperature range except that the modes related to tetragonal structure appear at low temperatures below 400 K in mixed-phase compositions (Raman spectra at a few temperatures for each composition of both the systems are shown in Figures S7-S14 in supplementary material). Phonon parameters are extracted at all temperatures by fitting the spectra with multiple Lorentzian functions. Most of the modes show a redshift with increasing temperature arising from the lattice anharmonicity. For example, the frequency of $E_{2g}$ [71 cm$^{-1}$] and $A_{1g}$ [104 cm$^{-1}$] phonons decreases with increasing temperature for all the compositions in both BaTi$_{1-x}$Mn$_x$O$_3$ and BaTi$_{1-x}$Fe$_x$O$_3$, as shown in Figures S15 and S16 in supplementary material. The frequency of $A_{1g}$ [600 cm$^{-1}$] phonon related to the displacements of magnetic (Fe) ions decreases with increasing temperature in the entire temperature range under investigation (80–800 K) (Figure 5). On the contrary, the modes at 152 cm$^{-1}$ ($E_{1g}$) and 635 cm$^{-1}$ ($A_{1g}$) exhibit anomalous behaviour (i.e. an increase in the frequency with increasing temperature) in both BaTi$_{1-x}$Mn$_x$O$_3$ and BaTi$_{1-x}$Fe$_x$O$_3$ compounds [Figure 4 and 5]. Importantly, we found no signatures of structural phase transition or change in the crystal symmetry in the investigated temperature range based on the Raman and x-ray diffraction measurements as a function of temperature (temperature-dependent PXRD is performed for BaTi$_{1-x}$Mn$_x$O$_3$: x = 0.02 and 0.3 and BaTi$_{1-x}$Fe$_x$O$_3$: x = 0.125 and 0.15 as shown in Figures S17 and S18 in supplementary material). The origin of observed phonon anomalies is analysed based on the structural as well as magnetic properties of the compounds and are discussed below.

In general, the temperature-dependent frequency of a phonon can be written as [54,55]:

$$\omega(T) = \omega_{anh}(T) + \Delta\omega_{el-ph}(T) + \Delta\omega_{sp-ph}(T) \quad (1)$$

The first term $\omega_{anh}(T)$ represents the frequency due to the anharmonicity which includes the quasi-harmonic contribution arising from the change in lattice volume and the phonon-phonon intrinsic anharmonic interactions. In the cubic anharmonic process, a phonon of frequency $\omega_0$ is assumed to decay into two phonons of equal frequency $\omega_0/2$ satisfying the energy and momentum conservation [55]. The contribution of the electron-phonon coupling, if any, to the phonon frequency is given by the term $\Delta\omega_{el-ph}(T)$ which is absent in the compounds studied here due to their electrically insulating nature. Finally, the renormalization of the phonon frequency due to spin-phonon coupling, if present, is accounted by the term $\Delta\omega_{sp-ph}(T)$. The temperature-dependent phonon frequency due to cubic anharmonicity (three-phonon process) can be written as [56,57]:

$$\omega_{anh}(T) = \omega_0 + A\left[1 + \frac{2}{\left(e^{\frac{\hbar\omega_0}{2k_BT}}-1\right)}\right] \quad (2)$$

Where $\omega_0$ is the frequency of the phonon at absolute zero temperature, $A$ is the coefficient of cubic anharmonicity, $\hbar$ is the reduced Planck constant, $k_B$ is the Boltzmann constant, and T is the variable temperature. The temperature-dependence of the modes at 71 cm$^{-1}$ ($E_{2g}$), 104 cm$^{-1}$ ($A_{1g}$) and 600 cm$^{-1}$ ($A_{1g}$) can be explained by Eq. 2 (see Figure S14 and S15 in supplementary material). The anomalous behaviour of the modes at 152 cm$^{-1}$ ($E_{1g}$) and 635 cm$^{-1}$ ($A_{1g}$), that respectively involve Ba and Ti atoms, can be attributed to strong anharmonicity arising from unusually large displacements of these atoms during their vibrations. The phonons involving Ba-O bonds may have higher anharmonicity relative to phonons associated with M-O bonds because of large Ba-O bond lengths (Figure S4-S6 and section S1 in supplementary material).

In order to investigate the possible association of the phonons with the magnetic properties of the compounds, we have measured the bulk magnetization with varying temperature and magnetic field. Figure 6 shows the magnetization as a function of temperature (MT) in the range of 2-300 K and magnetic field (MH) in the range of $\pm 7$ T for $BaTi_{1-x}Mn_xO_3$. No signatures of magnetic transition are observed in MT curves and even in $\frac{dM}{dT}$ $vs$ $T$ curves (not shown here) in the temperature range from 80-300 K. The MH curves are linear for all compositions which imply that the $BaTi_{1-x}Mn_xO_3$ system is in the paramagnetic phase. Besides that, a low-temperature transition at T*~ 43 K is observed in $BaTi_{1-x}Mn_xO_3$ for x $\geq$ 0.15 which is associated with either ferrimagnetic transition of Mn related phases or spin canting [58-60]. To be noted that this transition has no effect on the phonons of $BaTi_{1-x}(Mn/Fe)_xO_3$ studied above 80 K. The temperature-dependence of magnetization of $BaTi_{1-x}Fe_xO_3$ from (a) 2-300 K, and (b) 300-800 K are shown in Figure 7(a) and 7(b), respectively, while the temperature derivative for all the MT curves are shown in the respective insets for a better determination of the transition temperatures. Signature of para- to ferro- magnetic transition is seen at $T_{C1}$ ~ 690 K in all the compositions, matching with the earlier reports [61, 34]. The ferromagnetic transition (at $T_{C1}$) is prominent for x = 0.125 to 0.2 and is very weak in x = 0.05, 0.10, and 0.25. The normalized magnetic moment (per unit gram) exhibits an increasing trend with increasing doping up to x = 0.15 that decreases upon further doping (above x = 0.15), similar to the trend seen in the lattice parameter, $a_{6H}$ $vs$ $x$ (Figure 2d). The obtained Curie-Weiss temperature ($\theta_C$) is –ve for x = 0.05, although it is small which indicates the presence of an antiferromagnetic ordering but is very weak due to large separation between Fe atoms at low concentrations. The $\theta_C$ for x > 0.05 becomes +ve and increases with doping (Figure S18 and S19 in supplementary material). This indicates a dominant ferromagnetic contribution for compositions with x > 0.05, where the strength of ferromagnetic order increases till x = 0.15 and then decreases with further doping. It should be noted here that the in-plane lattice parameter, magnetic moment, and paramagnetic Curie temperature ($\theta_C$) show similar trend with Fe-doping suggesting a strong correlation between structural and magnetic degrees of freedom mediated through the lattice strain at the boundary of the two phases in mixed tetragonal and 6H-compositions.

As the temperature is lowered further, magnetic moment shows another upturn at $T_{C2} \sim 407$ K for $x \leq 0.15$ and the corresponding signature is seen in the temperature-derivative curves. Notably, $T_{C2}$ is observed in the compositions having perovskite-type (tetragonal) phase fraction. Therefore, its origin can be attributed to the paraelectric to ferroelectric (accompanied by cubic to tetragonal) phase transition in the mixed phase compositions thus indicating the presence of strain-induced magnetoelectric coupling [62, 63]. When the perovskite unit cells undergo phase transition from cubic to tetragonal at $T_{C2}$, the induced strain on 6H unit cells modifies the exchange interactions (Fe-O-Fe bond angles and bond lengths) and may give rise to an anomaly in magnetization. In addition to this, for the compositions with $x = 0.20$ and $0.30$, that stabilizes in single 6H phase, exhibit a weak change in magnetization (as seen in dM/dT) at $T_{C3} \sim 110$ K having a possible association with either the paraelectric-ferroelectric (at $\sim 74$ K for 6H-BaTiO$_3$ [10]) or ferro-to-antiferromagnetic transition (at $\sim 130$ K for 6H-BaFeO$_{3-\delta}$ [29]) of the 6H unit cells. Latter one can be ruled out as the MH curve measured at 5 K has significant loop-opening (Figure S21 in supplementary material). We believe that the origin of $T_{C3}$ may be related to the para-to-ferroelectric transition of the 6H phase.

The field-dependent magnetization (MH) measured at room temperature (Figure 8a) is linear for $x = 0.05$ while it is non-linear for all the other compositions of BaTi$_{1-x}$Fe$_x$O$_3$ thus implying that the system is ferromagnetic for $x > 0.05$. The pinched hysteresis (MH) loops suggest that the magnetic ground state has competing ferromagnetic and antiferromagnetic interactions [64]. The presence of magnetic anomaly at ferroelectric transition in MT indicates the presence of magnetoelectric coupling in BaTi$_{1-x}$Fe$_x$O$_3$: $x = 0.1$ to $0.15$. Figure 8(b) shows the coercivity as a function of Fe concentration at a few temperatures. The MH curves at a few temperatures are plotted in Figure S21 in supplementary material. It is very clear that the coercive field increases with increasing temperature which corroborates the magnetoelectric coupling in BaTi$_{1-x}$Fe$_x$O$_3$ system [65, 66]. It should be noted here that the 6H structure consists of dimers of Fe$_2$O$_9$ of face-shared octahedra with a significantly short Fe-Fe distance apart from corner-shared octahedra. The mixed-valence state of Fe ions along with the structural peculiarity leads to the coexistence of various competing magnetic interactions in this phase. Notably, our observation of ferromagnetism is in agreement with a theoretical prediction by Xu et al. [17] where they have shown that the ferromagnetic (FM) configuration has lower energy than the antiferromagnetic (AFM) configuration for BaTi$_{0.875}$Fe$_{0.125}$O$_3$. The transition temperature $T_{C1}$ observed in our data also matches well with their report [17]. In the stoichiometric scenario with all $Fe^{+4}$ ions, spins are expected to be aligned antiparallel to each other along the *c*-axis (parallel in the plane with no available direct exchange paths). In this configuration, there exists two pathways for the superexchange interaction: one is M1-O2-M2 ($Fe^{+4}$-O2-$Fe^{+4}$: θ ~ 177°) across the corner shared and face-shared octahedra which leads to antiferromagnetic ordering and the other one is M2-O1-M2 ($Fe^{+4}$-O1-$Fe^{+4}$: θ ~ 85°) across two face-shared octahedra leading to weak ferromagnetic ordering (dependence of these on doping is shown in Figure S22). The origin of FM in Fe-doped $BaTiO_3$ is addressed by several authors [17, 21, 30, 33, 35-37, 61] where the possible mechanisms include superexchange, double exchange, RKKY interactions, cation ordering, lattice strain, and the formation of bound magnetic polarons. We rule out the RKKY interaction in our samples due to their insulating nature. Since sintering temperature during synthesis is less than 1500°C, cation ordering is less likely to be present or minimal [37]. The lattice strain at the boundary of two types of unit cells decreases with Fe concentration due to a decrease in the phase fraction of perovskite-type unit cells which would eventually cause a decrease in the magnetic parameters (magnetic moment, coercive field, and Curie-Weiss temperature) too. The otherwise observed behaviour suggests that the strain-induced FM is not sufficient to explain our results. Since oxygen vacancies are known to be present in this structure and the presence of $Fe^{+3}$ ions along with $Fe^{+4}$ ions are evidenced by PXRD, double exchange $Fe^{+3}$-O-$Fe^{+4}$ interaction is also possible giving rise to FM. The increment in the ferromagnetic parameters with doping in the range of x = 0.05 to 0.15 is similar to the case of diluted magnetic semiconductors where bound magnetic polarons (BMP) are responsible for ferromagnetism [35]. We believe that the magnetism observed in our samples arises from superexchange, double exchange, and the formation of BMPs.

As we have seen from our magnetic studies, the $BaTi_{1-x}Mn_xO_3$ is paramagnetic and $BaTi_{1-x}Fe_xO_3$ is composition-dependent magnetic with competing ferro-and antiferro-magnetic interactions. However, phonon anomalies are present in both the systems and surprisingly, the anomaly decreases with dopant (Mn/Fe) concentration in both the compounds irrespective of the differences in their magnetic ground state and temperature-dependence of the lattice parameters (unit cell dimensions and volume). Therefore, the anomalous temperature-dependence of the $E_{1g}$ at 152 cm$^{-1}$ phonon in $BaTi_{0.98}Mn_{0.02}O_3$ and $BaTi_{0.95}Fe_{0.05}O_3$ can be attributed to a strong anharmonicity arising from large displacement of the Ba-atoms associated with long Ba-O bond length (see section S1 and Figure S6 in supplementary material for a discussion). Similarly, the anomaly in $A_{1g}$ mode (at ~ 635 cm$^{-1}$ in $BaTi_{1-}$ $_x$Mn$_x$O$_3$ as well as at ~ 637 cm$^{-1}$ in BaTi$_{1-x}$Fe$_x$O$_3$) also can be attributed to strong anharmonicity arising from large displacements of O atom (in Ti-O bond) during the vibrations. As the doping increases, the Ba-O/Ti-O bond length decreases thus reducing the displacements during the vibration and in turn reducing the phonon anomaly/anharmonicity.

In addition to the anomalous temperature-dependence of phonon frequency ($\omega$ vs $T$), the clear slope change around T$_{C2}$ ~ 407 K is observed for $E_{1g}$ [152 cm$^{-1}$] and $A_{1g}$ [635 cm$^{-1}$] modes in BaTi$_{0.98}$Mn$_{0.02}$O$_3$ and BaTi$_{1-x}$Fe$_x$O$_3$: x = 0.05 to 0.15. To recall, these compositions have the perovskite-type BaTiO$_3$ phase fraction which undergoes a transition from high-temperature cubic to low-temperature tetragonal phase at ~ 400 K introducing adistortion in the TiO$_6$ octahedra at low temperatures. This distortion in the perovskite unit cells strains the 6H unit cells in the mixed phase compositions thus influencing the phonon behaviour across the phase transition exhibiting a change in the slope.

**Conclusions**

We have investigated the correlation between phonons and multiferroic order parameters in BaTi$_{1-x}$M$_x$O$_3$: M = Mn, Fe systems using powder x-ray diffraction (PXRD), Raman spectroscopic, and magnetic measurements. Anomalous temperature-dependence is observed for two phonons associated with the displacements of Ba ions ($E_{1g}$ at ~ 152 cm$^{-1}$) and O ions located in TiO$_6$ octahedra ($A_{1g}$ at ~ 636 cm$^{-1}$), which get suppressed by doping (Mn/Fe). Bulk magnetization measurements reveal the existence of three magnetic transitions T$_{C1}$ ~ 680 K, T$_{C2}$ ~ 407 K, and T$_{C3}$ ~ 110 K: where T$_{C1}$ corresponds to a ferromagnetic transition, while the origin of T$_{C2}$ and T$_{C3}$ are associated with the ferroelectricity of tetragonal (perovskite) and 6H phases, respectively, in BaTi$_{1-x}$Fe$_x$O$_3$, while, BaTi$_{1-x}$Mn$_x$O$_3$ is paramagnetic in the entire investigated composition and temperature ranges. Phonon anomalies are observed to be independent of magnetic ordering and, hence, can be attributed to strong anharmonicity arising from the large displacements of the atoms involved. The response of magnetism across the para-to-ferroelectric transitions and the corresponding response of coercive field with temperature indicate the presence of magnetoelectric coupling in the BaTi$_{1-x}$Fe$_x$O$_3$ compounds. Additional evidence of the coupling is seen as a change in the phonon behaviour across the ferroelectric transition temperature (T$_{C2}$) in BaTi$_{1-x}$Fe$_x$O$_3$ and BaTi$_{0.98}$Mn$_{0.02}$O$_3$ in the mixed phase compositions. We believe that our results will excite further theoretical and experimental studies on these compounds with mixed compositions with potential technological applications.

**Acknowledgement**

Authors acknowledge IISER Bhopal for research facilities, B. P. acknowledges the University Grant Commission for fellowship and S. S. acknowledges DST/SERB (project Nos. ECR/2016/001376 and CRG/2019/002668) and Nano-mission (Project No. SR/NM/NS-84/2016(C)) for research funding. Support from DST-FIST (Project No. SR/FST/PSI-195/2014(C) is also thankfully acknowledged.

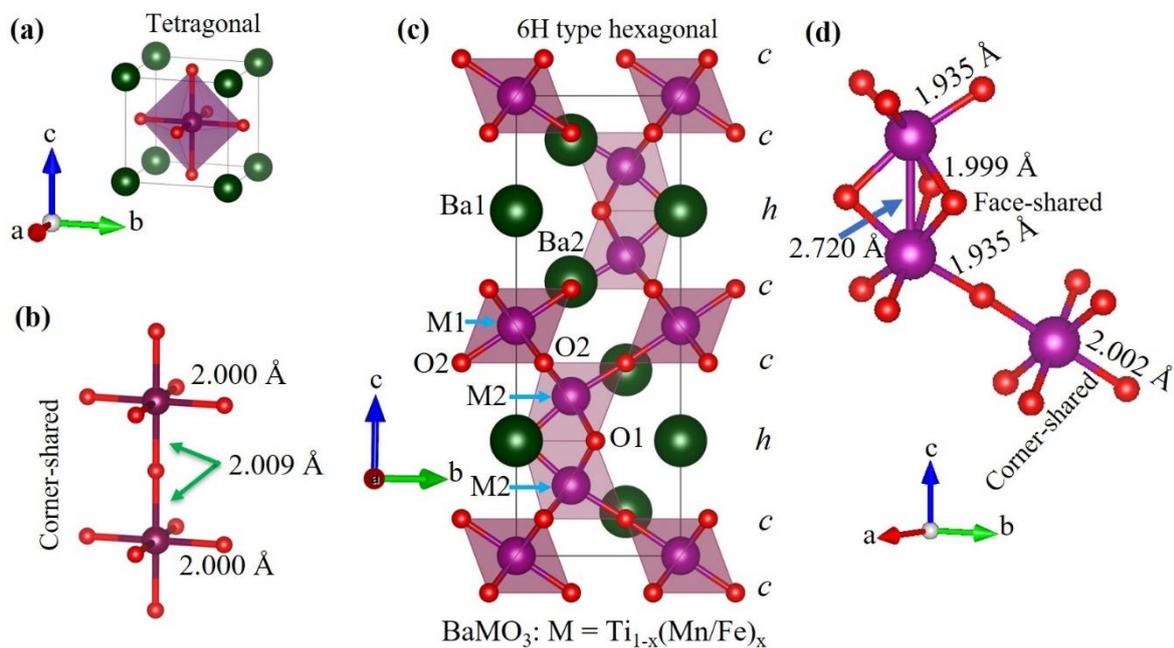

Figure 1. Crystal structure and corresponding bond lengths in tetragonal (a, b) and 6H phases (c, d) of BaMO$_3$: M = Ti$_{1-x}$(Mn/Fe)$_x$ at room temperature.

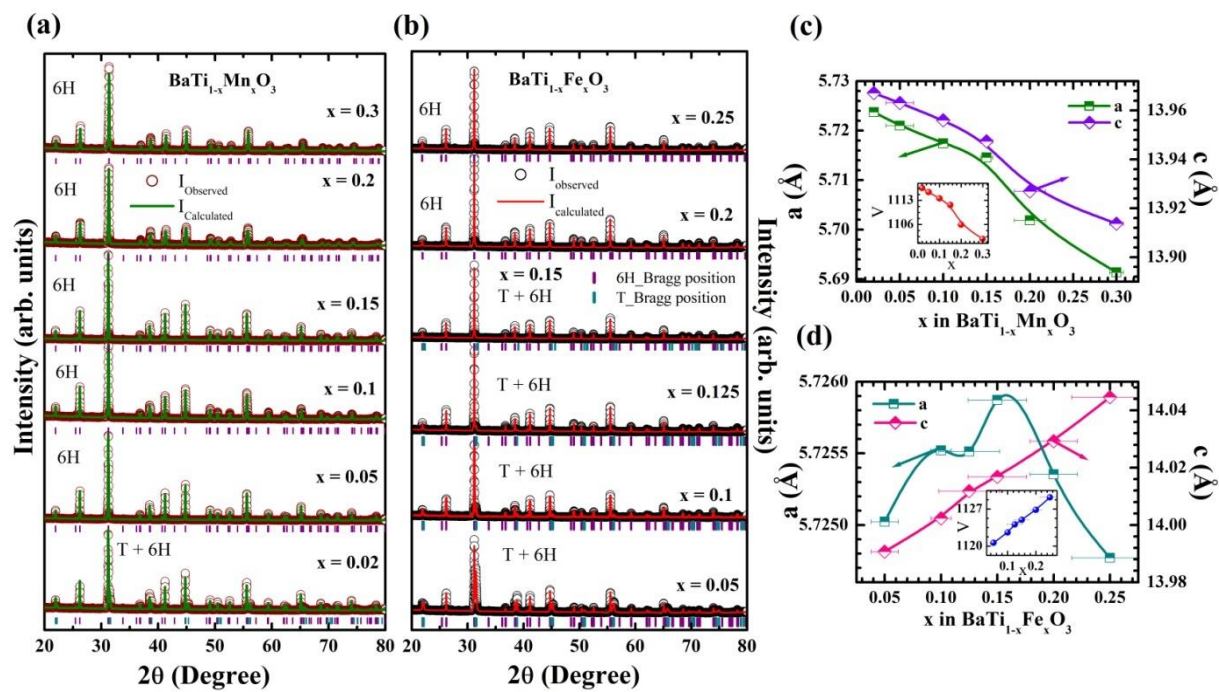

Figure 2. X-ray diffraction patterns for (a) $BaTi_{1-x}Mn_xO_3$ and (b) $BaTi_{1-x}Fe_xO_3$ collected at room temperature, (c) and (d) show the lattice parameters as a function of doping (Mn/Fe), respectively. Insets show the respective unit cell volume with doping (Mn/Fe).

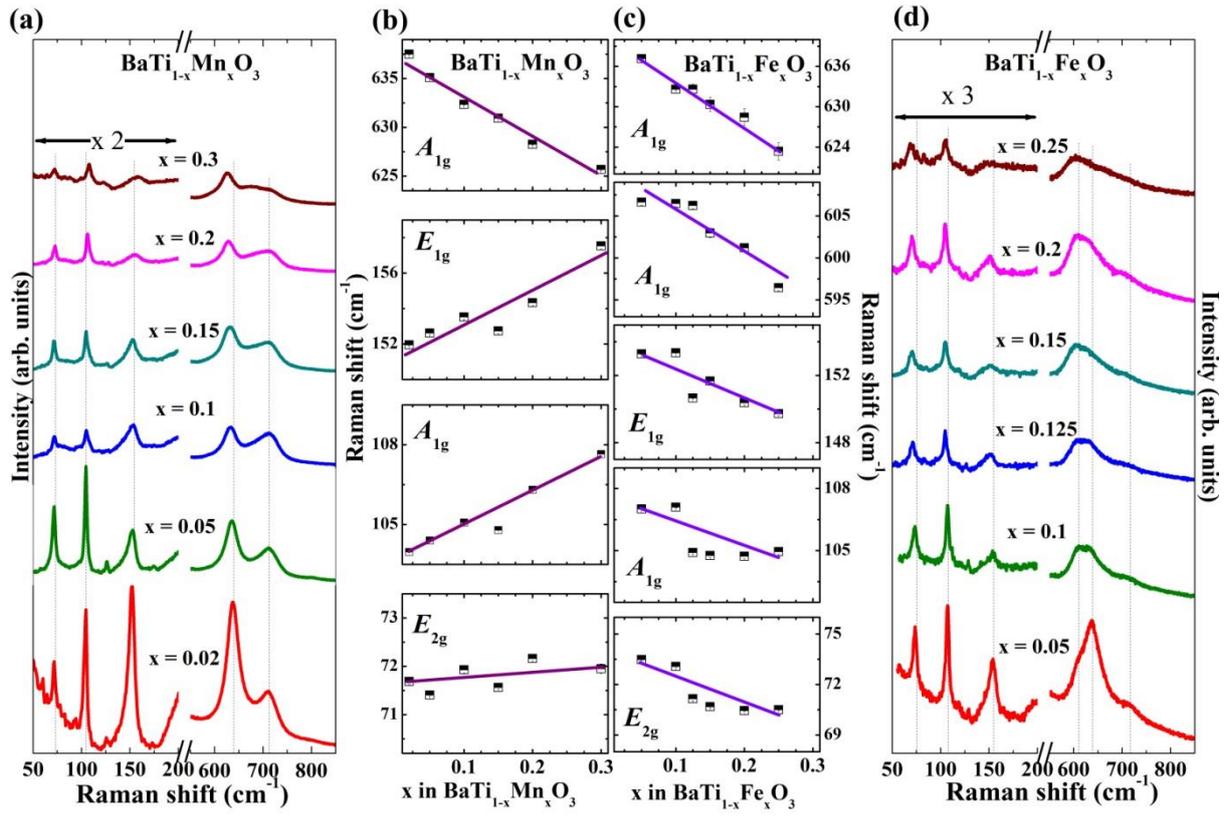

Figure 3. Raman spectra and phonon frequency as a function of (Mn/Fe) doping for BaTi$_{1-x}$Mn$_x$O$_3$ (a, b) and BaTi$_{1-x}$Fe$_x$O$_3$ (c, d).

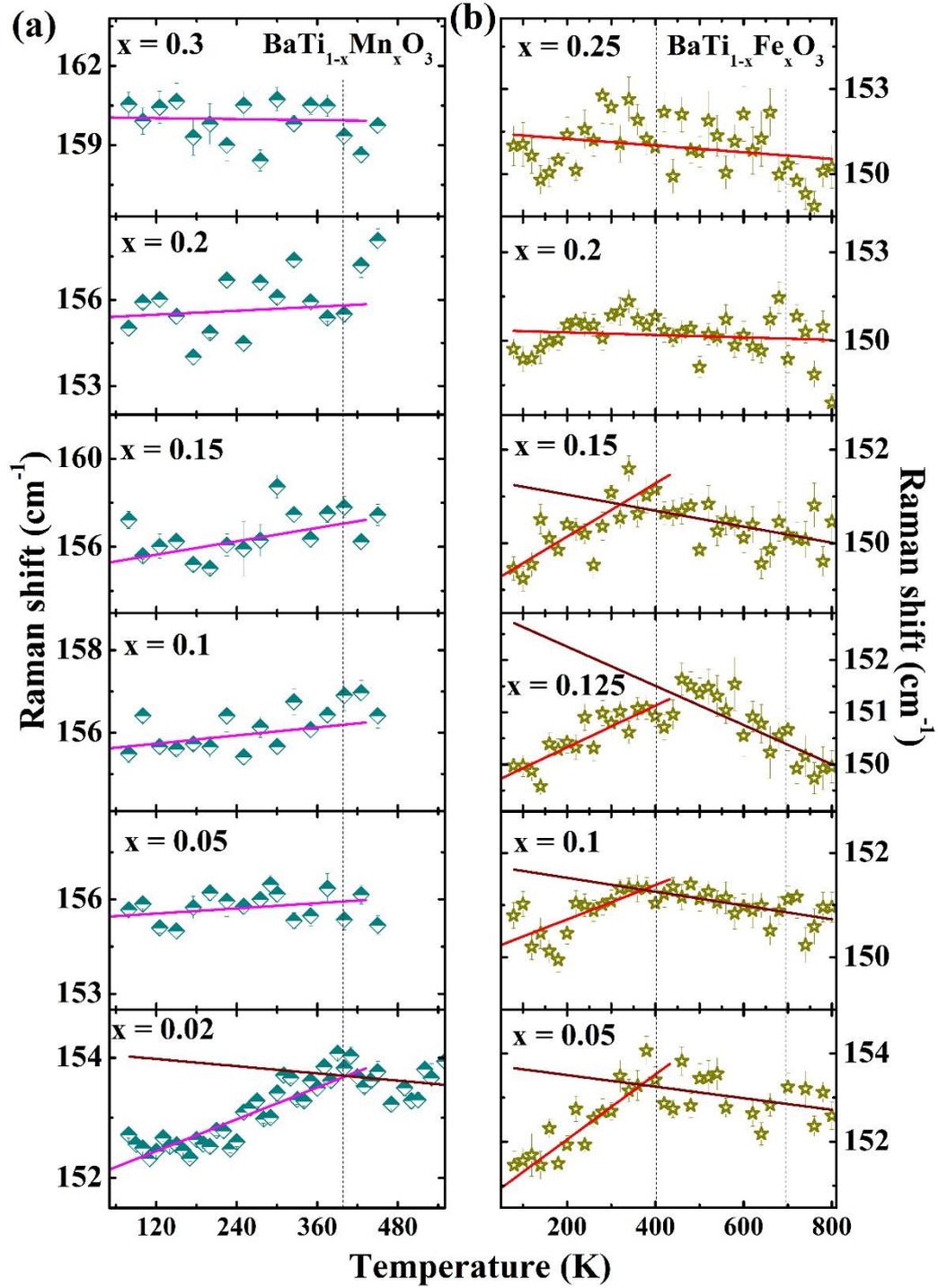

Figure 4. The phonon frequency as a function of temperature for $E_{1g}$ mode at 152 cm$^{-1}$ for all compositions with (a) Mn-doping and (b) Fe-doping. The modes respond to the cubic-to-tetragonal phase transition of BaTiO$_3$ at ~ 407 K as a change in slope. Further, the modes show anomalous behaviour with temperature that reduces with doping (discussed in text).

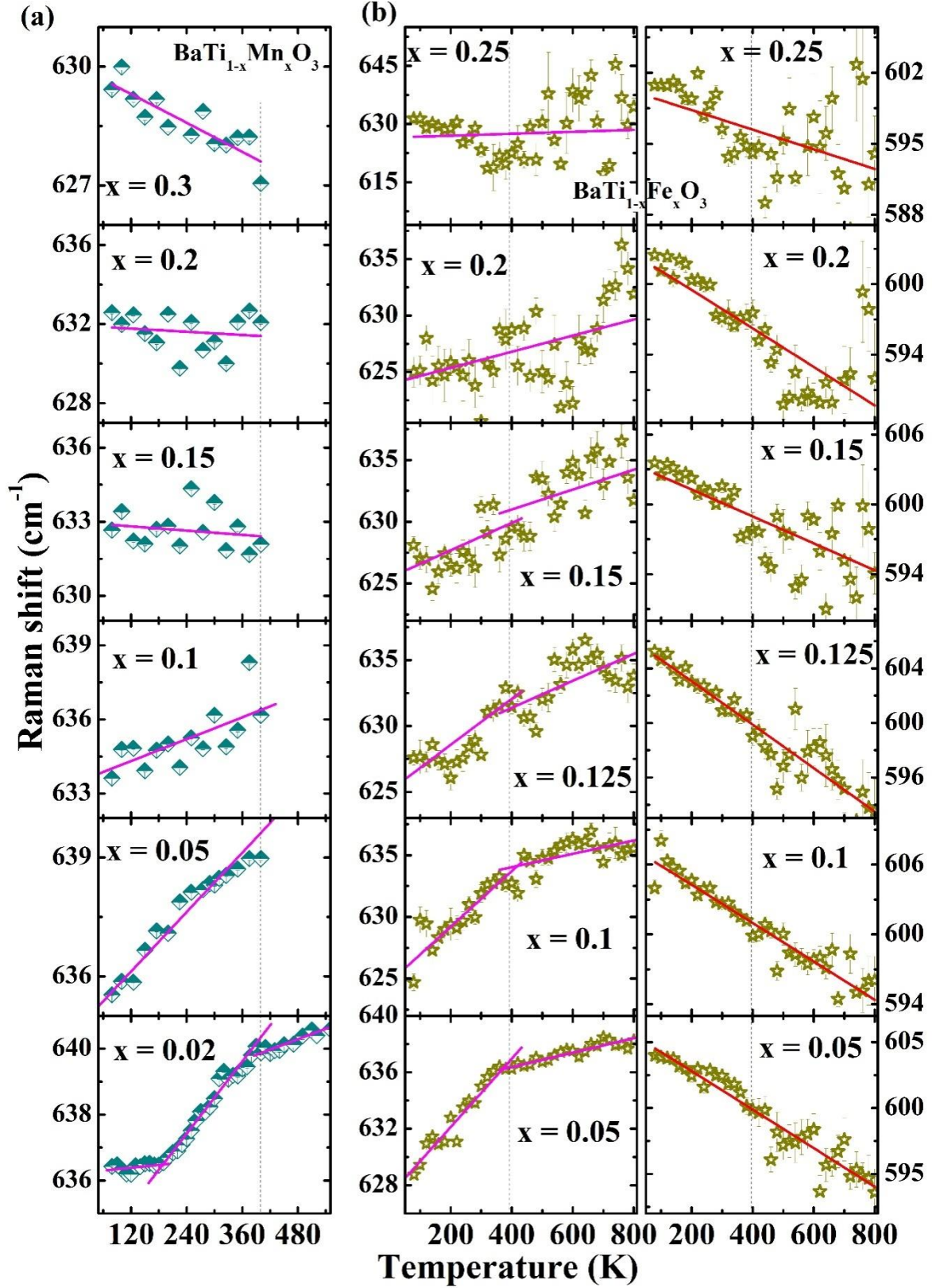

Figure 5. Temperature-dependence of the O-vibration of TiO$_6$ octahedra ($A_{1g}$) at ~ 636 cm$^{-1}$ in BaTi$_{1-x}$Mn$_x$O$_3$ (a), (b) Fe-doping in BaTiO$_3$ splits this mode into two (one at ~ 636 cm$^{-1}$ involving Ti and another at ~ 600 cm$^{-1}$ involving Fe) due to the mass difference of Ti and Fe.

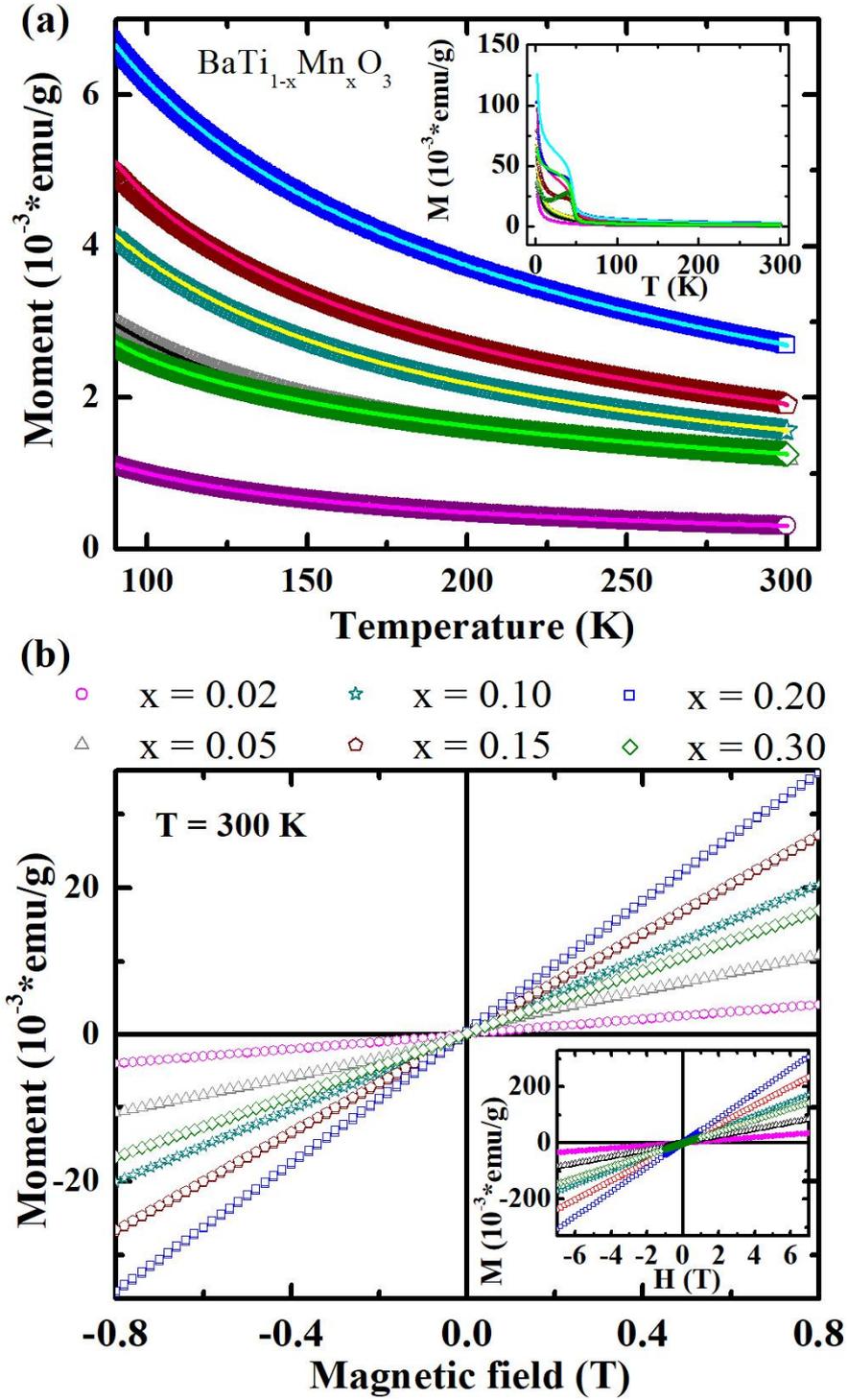

Figure 6. Magnetization as a function of (a) temperature (MT) and (b) magnetic field for BaTi$_{1-x}$Mn$_x$O$_3$. In (a) open symbols and solid lines represent ZFC and FC measurements of MT, respectively. MT in the range of 2-300 K and MH in the field range of ± 7T are shown in insets of (a) and (b), respectively.

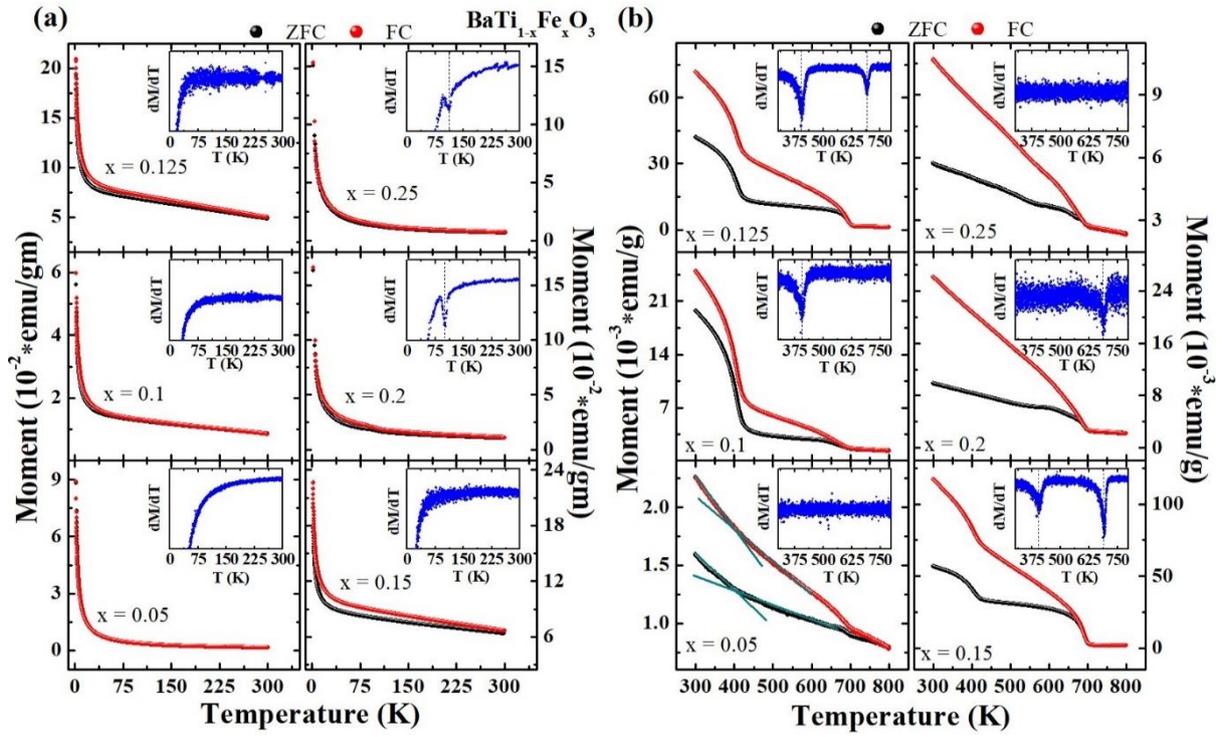

Figure 7. Magnetization as a function of temperature (MT) for BaTi$_{1-x}$Fe$_x$O$_3$. Insets show temperature derivative of the respective MT curve.

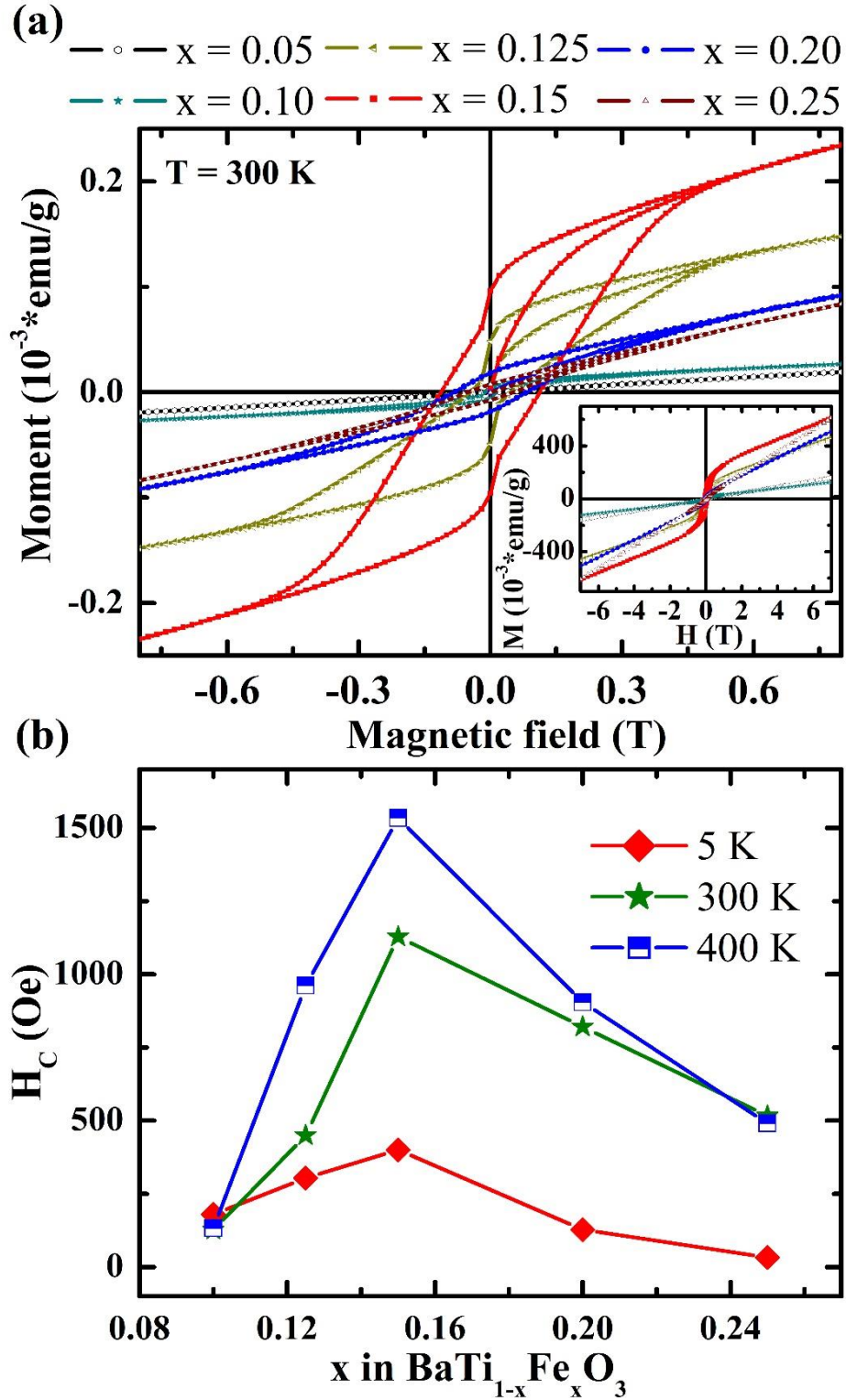

Figure 8. Magnetization as a function of (a) magnetic field and (b) coercive field as a function of doping (Fe) in BaTi$_{1-x}$Fe$_x$O$_3$.

# Supplementary material
# Correlations between the structural, magnetic, and ferroelectric properties of BaMO$_3$: M = Ti$_{1-x}$(Mn/Fe)$_x$ compounds: A Raman study

Bommareddy Poojitha, Ankit Kumar, Anjali Rathore, and Surajit Saha*

Department of Physics, Indian Institute of Science Education and Research Bhopal, 462066, India

This supplementary material contains additional data on x-ray diffraction, Raman spectroscopy, and magnetic measurements of BaMO$_3$: M = Ti$_{1-x}$(Mn/Fe)$_x$. The figures and respective details are given below.

## S1. X-ray diffraction

Phase fractions

The weight fractions of the respective crystal phases at each composition of $BaTi_{1-x}Mn_xO_3$ and $BaTi_{1-x}Fe_xO_3$ are calculated from Rietveld refinement (Table S1) using High-score Plus software based on the equation,

$$w_p = (SZMV)_p / \sum_i (SZMV)_i \quad\quad\quad\quad\quad (S1)$$

where, $w_p$ is weight fraction of phase $p$, $S$ is the scale factor, $Z$ is the number of formula units per unit cell, $V$ is unit cell volume, and $i$ is index running over all phases.

Table S1. Phase fractions of tetragonal and 6H structures in $BaTi_{1-x}Mn_xO_3$, and $BaTi_{1-x}Fe_xO_3$ systems.

| x in $BaTi_{1-x}Mn_xO_3$ | T (%) | 6H (%) | x in $BaTi_{1-x}Fe_xO_3$ | T (%) | 6H (%) |
|---|---|---|---|---|---|
| x = 0.02 | 18.8 | 81.2 | x = 0.05 | 47.5 | 52.5 |
| x = 0.05 | -- | 100 | x = 0.1 | 16 | 84 |
| x = 0.1 | -- | 100 | x = 0.125 | 10.5 | 89.5 |
| x = 0.15 | -- | 100 | x = 0.15 | 1.3 | 98.7 |
| x = 0.2 | -- | 100 | x = 0.2 | 0 | 100 |
| x = 0.3 | -- | 100 | x = 0.25 | 0 | 100 |

# Lattice parameters

Figure S1 shows the lattice parameters of BaTi$_{1-x}$Fe$_x$O$_3$ corresponding to tetragonal and 6H phase. In-plane and out of plane lattice parameters corresponding to tetragonal phase increase and decrease with doping (Fe), respectively. On the other hand, in-plane lattice parameter corresponding to 6H phase increases with doping till x = 0.15 and then decreases with further doping and out of plane lattice parameter increases with doping (Fe).

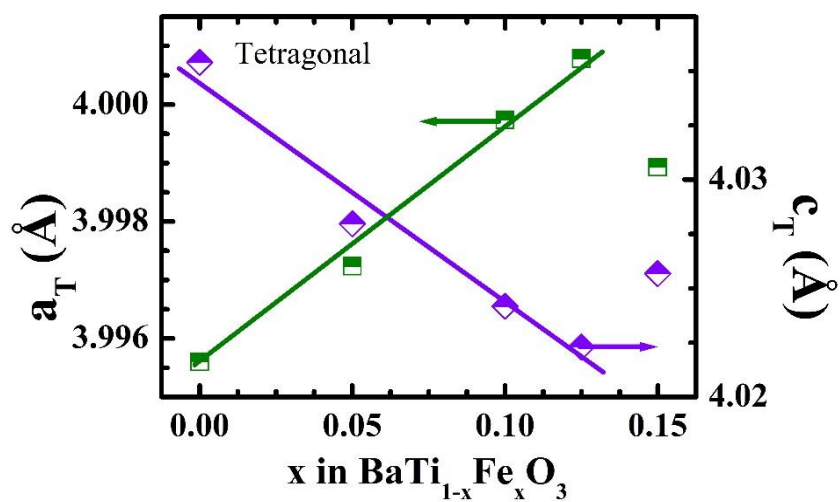

Figure S1. Lattice parameters of BaTi$_{1-x}$Fe$_x$O$_3$ corresponding to tetragonal phase.

## S2. Composition analysis: EDAX

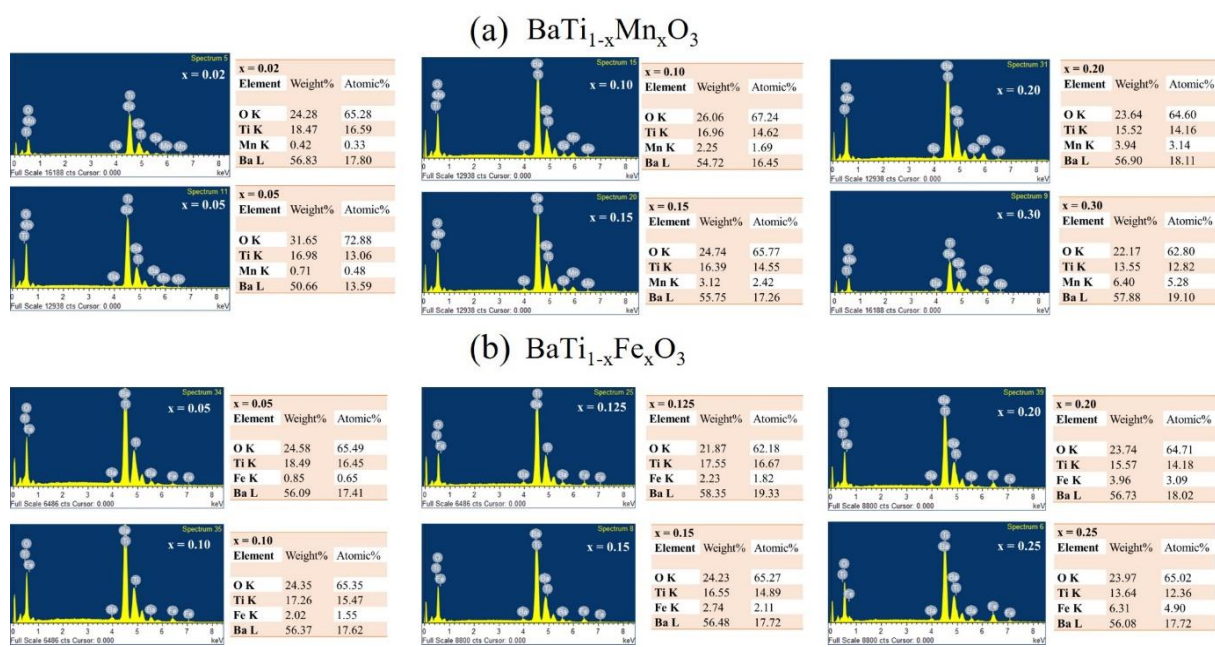

Figure S2. EDAX spectra for $BaTi_{1-x}Mn_xO_3$ and $BaTi_{1-x}Fe_xO_3$

EDAX spectra (Figure S2) confirm that the chemical compositions agree well with the expected stoichiometry, as shown in Table S2. The error bars for each composition are extracted and plotted in Figure 2 in main text.

Table S2: EDAX data of $BaTi_{1-x}Fe_xO_3$ and $BaTi_{1-x}Mn_xO_3$

| | $BaTi_{1-x}Fe_xO_3$ | | | | | |
|---|---|---|---|---|---|---|
| **X (Planned)** | 0.05 | 0.10 | 0.125 | 0.15 | 0.20 | 0.25 |
| **Element** | Atomic% | Atomic% | Atomic% | Atomic% | Atomic% | Atomic% |
| **O K** | 65.49 | 65.35 | 62.18 | 65.27 | 64.71 | 65.02 |
| **Ti K** | 16.45 | 15.47 | 16.67 | 14.89 | 14.18 | 12.36 |
| **Fe K** | 0.65 | 1.55 | 1.82 | 2.11 | 3.09 | 4.90 |
| **Ba L** | 17.41 | 17.62 | 19.33 | 17.72 | 18.02 | 17.72 |
| **x (Obtained from EDAX)** | 0.038 | 0.091 | 0.098 | 0.124 | 0.179 | 0.284 |
| | $BaTi_{1-x}Mn_xO_3$ | | | | | |
| **X (Planned)** | 0.02 | 0.05 | 0.10 | 0.15 | 0.20 | 0.30 |
| **Element** | Atomic% | Atomic% | Atomic% | Atomic% | Atomic% | Atomic% |
| **O K** | 65.28 | 72.88 | 67.24 | 65.77 | 64.60 | 62.80 |
| **Ti K** | 16.59 | 13.06 | 14.62 | 14.55 | 14.16 | 12.82 |
| **Mn K** | 0.33 | 0.48 | 1.69 | 2.42 | 3.14 | 5.28 |
| **Ba L** | 17.80 | 13.59 | 16.45 | 17.26 | 18.11 | 19.10 |
| **x (Obtained from EDAX)** | 0.019 | 0.034 | 0.103 | 0.143 | 0.182 | 0.292 |

## S3. Effect of doping on phonons at room temperature

Figure S3 shows the Raman spectra at room temperature. The ionic radii are $Ti^{+4}$ (0.605 Å), $Mn^{+4}$ (0.530 Å), $Fe^{+4}$ (0.585 Å), and $Fe^{+3}$ (0.645 Å). The atomic masses of Ti, Mn, and Fe are 47.68, 54.94, and 55.85, respectively. The phonon frequency ($\omega$) under harmonic approximation can be written as $\omega \propto \sqrt{\frac{k}{m}}$, where m is the effective mass of ions participating in vibration and k is the spring constant which is related to the strength of chemical bonds. The frequency of mode at 630 cm$^{-1}$ in BaTi$_{1-x}$Mn$_x$O$_3$ is decreases with increasing Mn content due to the fact that Mn is heavier than Ti. On the other hand, the frequency of modes at 72 and 107 cm$^{-1}$ increases with doping which is due to the effect of change in ionic radius. The frequency of all phonons in BaTi$_{1-x}$Fe$_x$O$_3$ are observed to be decreasing with increasing Fe content which can be attributed to the effect of increase in mass upon incorporating Fe in place of Ti. To be noted that the peak around 308 cm$^{-1}$ marked in Figure S3 indicates the presence of tetragonal (perovskite) phase [1-5]. The additional peak appears around 714 cm$^{-1}$ which is not expected for the 6H phase. It may be associated with tetragonal unit cells that are plausibly present in the sample at extremely low concentration that XRD could not detect in case of BaTi$_{1-x}$Mn$_x$O$_3$: x > 0.02 and BaTi$_{1-x}$Fe$_x$O$_3$: x = 0.20 and 0.25. Otherwise, it may be a defect induced Raman active phonon band.

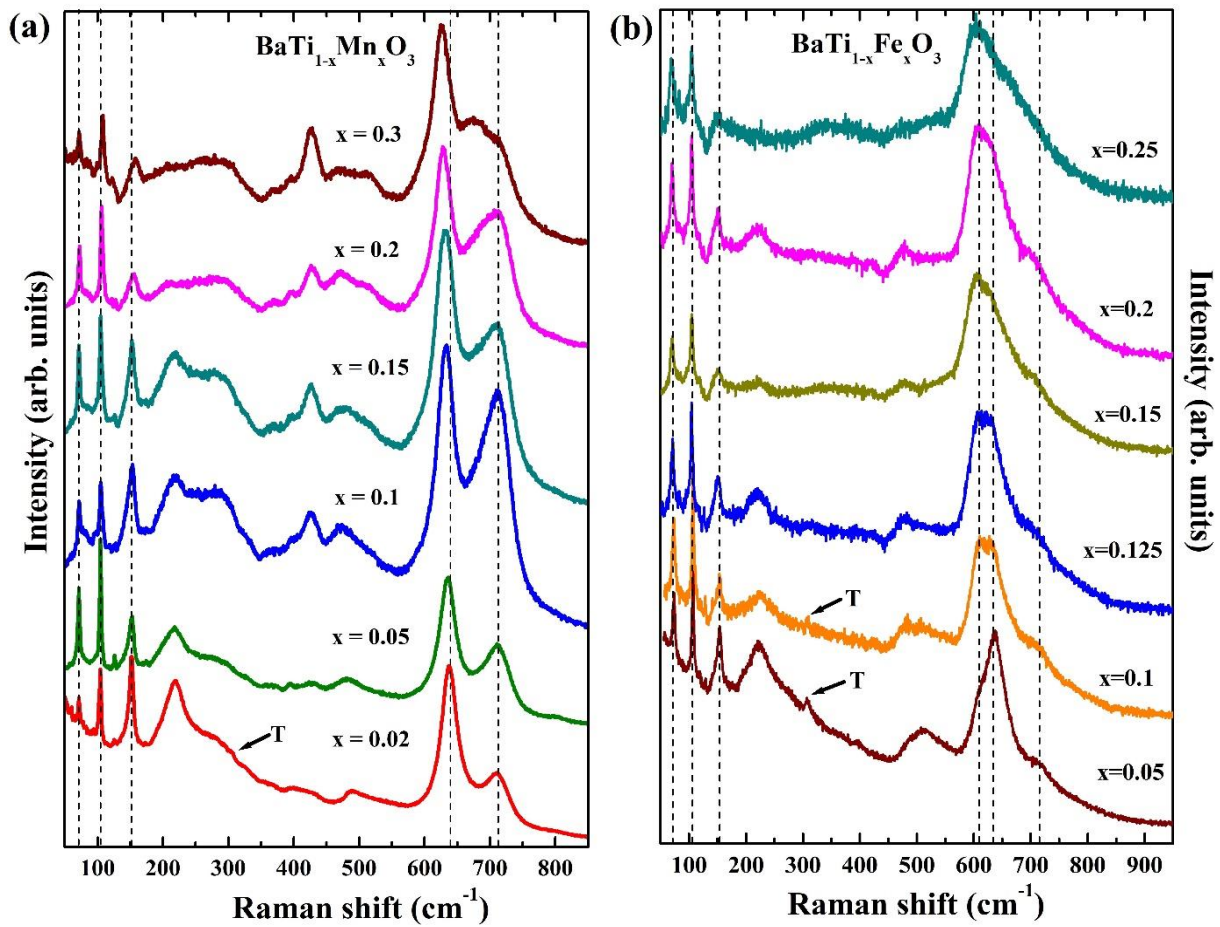

Figure S3. (a) Raman spectra of BaTi$_{1-x}$Mn$_x$O$_3$: x = 0.02 to 0.3 and (b) BaTi$_{1-x}$Fe$_x$O$_3$: x = 0.05 to 0.25, respectively at room temperature. A weak and sharp peak can be noticed at ~ 308 cm$^{-1}$ which is a signature Raman mode of tetragonal BaTiO$_3$.

**S4. Bond lengths and phonon anharmonicity**

The bond lengths for Ba-O and M-O bonds are displayed in Figure S4 and S5, found to be decreasing in BaTi$_{1-x}$Mn$_x$O$_3$ and increasing in BaTi$_{1-x}$Fe$_x$O$_3$, respectively, with doping (Mn/Fe). The bond lengths can be directly connected to the anharmonicity of the crystal through a simple anharmonic potential. To consider a simple example of an anharmonic potential, such as, Lenard-Jones type potential, $V(r)$ is:

$$V(r) = 4\varepsilon \left[ \left(\frac{\sigma}{r}\right)^{12} - \left(\frac{\sigma}{r}\right)^{6} \right] \quad \ldots\ldots\ldots\ldots (S2)$$

where, $r$ is the relative displacement of an atom w.r.t. its equilibrium position (or another atom that it is bonded to), $\sigma$ is a parameter. As can be seen in Figure S6, the bottom of the potential is harmonic-type for small displacements ($r$) (Figure S6(a)). However, for large displacements the potential $V(r)$ tends to become more and more anharmonic (Figure S6(b)).

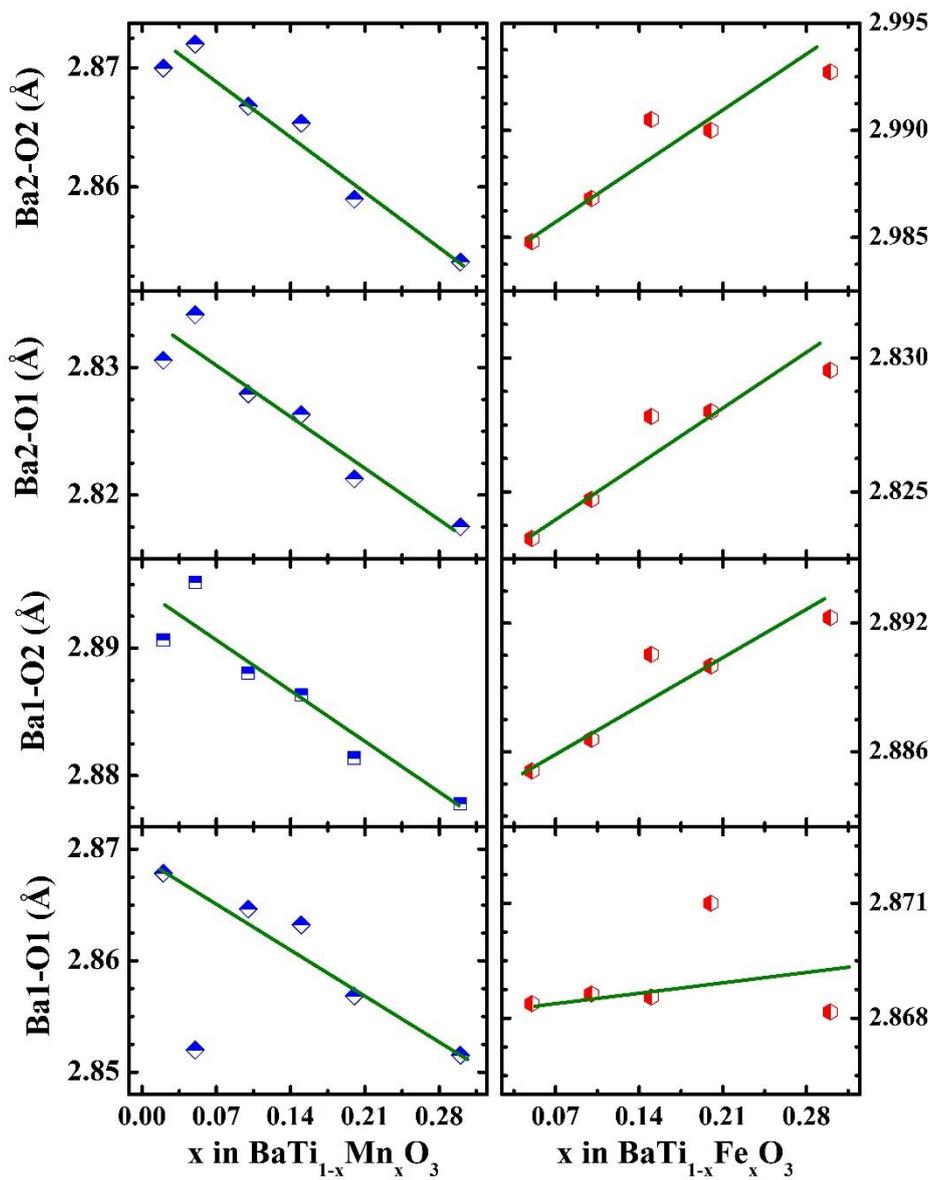

Figure S4. Ba-O bond lengths as a function of doping (Mn/Fe) in BaTi$_{1-x}$Mn$_x$O$_3$ and BaTi$_{1-x}$Fe$_x$O$_3$.

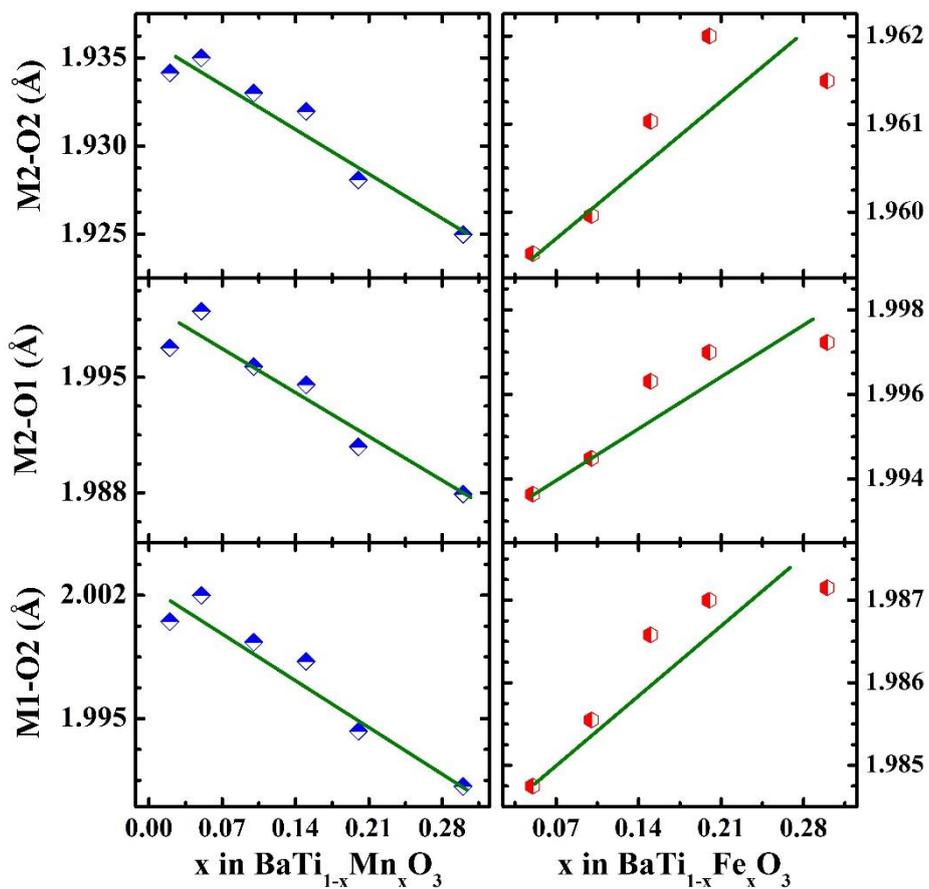

Figure S5. M-O (M = Mn/Fe) bond lengths as a function of doping (Mn/Fe) in $BaTi_{1-x}Mn_xO_3$ and $BaTi_{1-x}Fe_xO_3$.

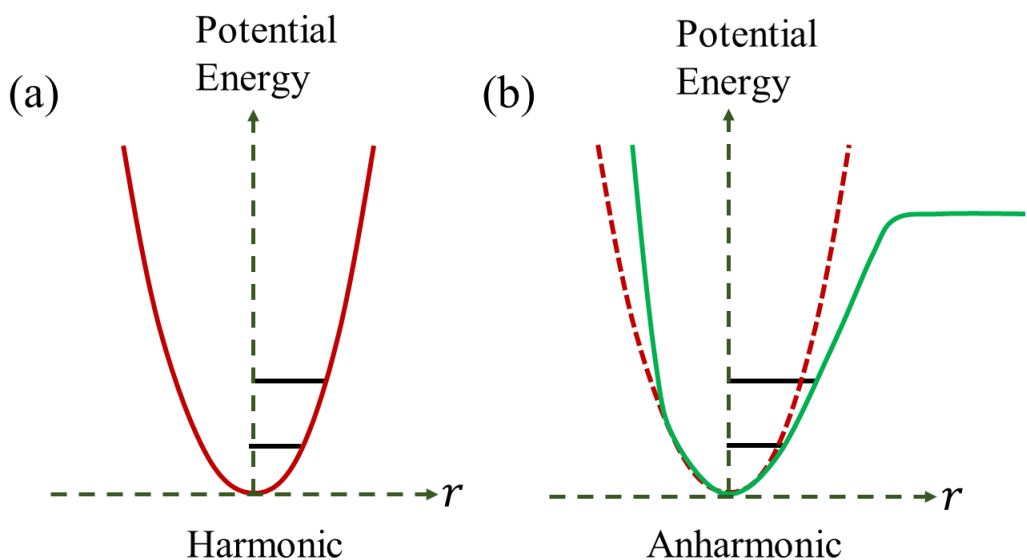

Figure S6. (a) Harmonic potential and (b) Anharmonic Lennard-Jones type potential (Eq. S2) as a function of atomic displacement.

## S5. Temperature dependence of Raman spectra

The Raman spectra at a few temperatures for each composition of both systems are displayed in Figure S7-14. The solid lines in Figure S15, S16 represent anharmonic fitting using Eq. 2 explained in main text.

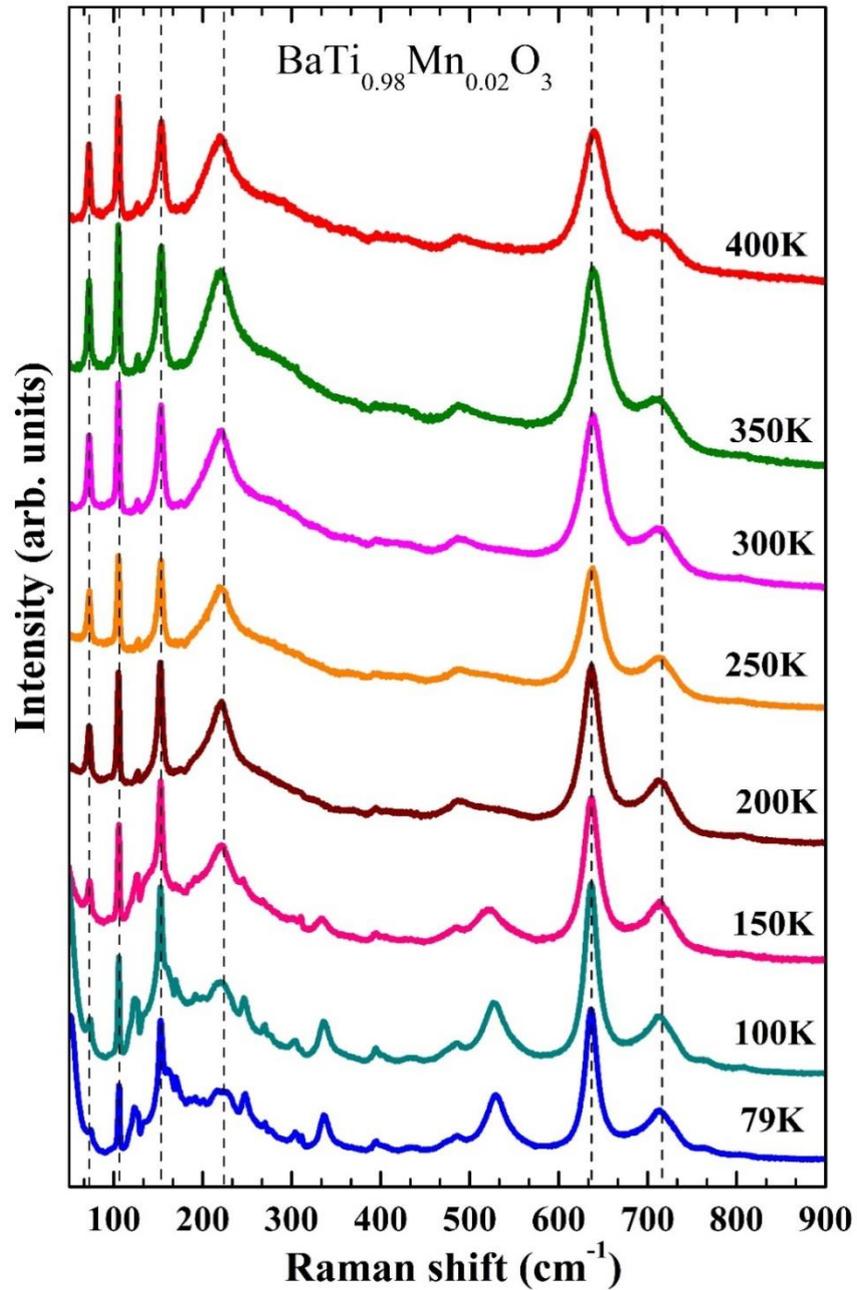

Figure S7. Raman spectra of BaTi$_{0.98}$Mn$_{0.02}$O$_3$ at a few typical temperatures.

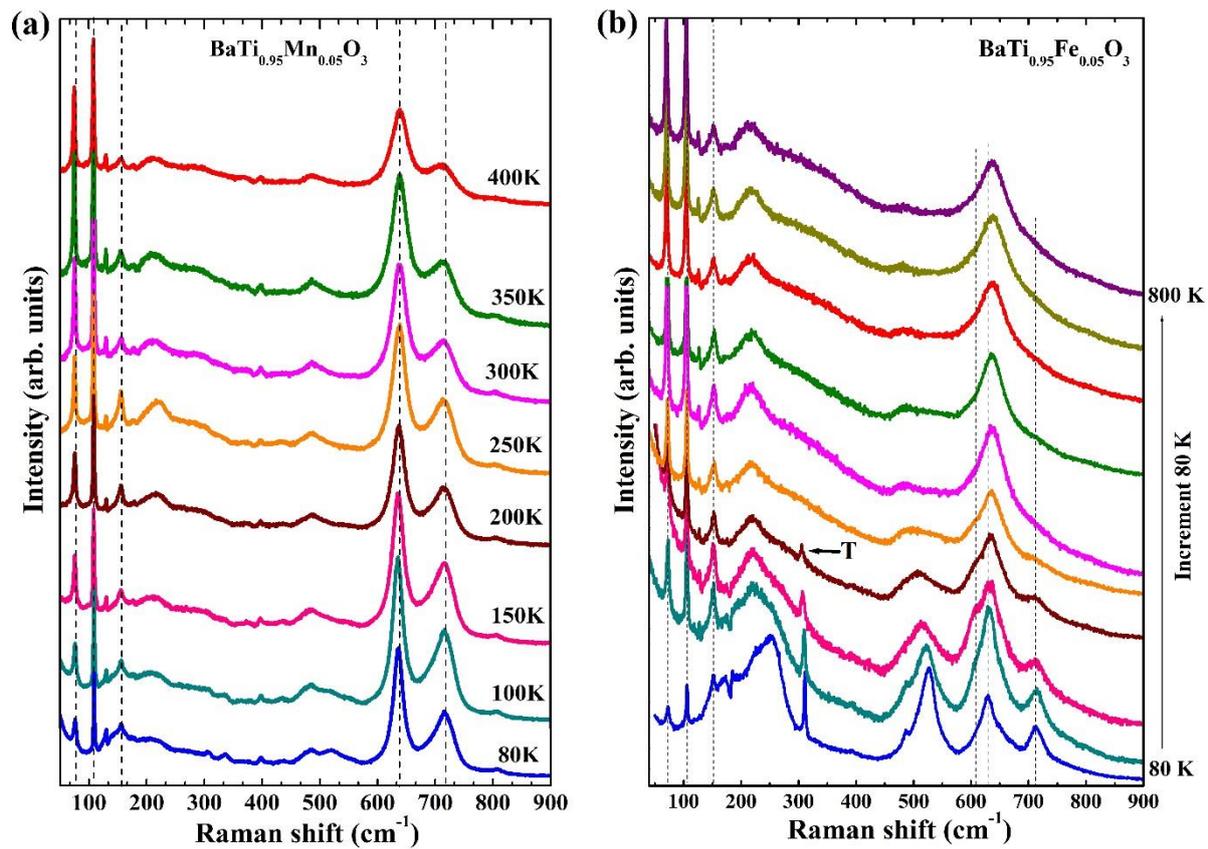

Figure S8. Raman spectra of BaTi$_{0.95}$Mn$_{0.05}$O$_3$ and BaTi$_{0.95}$Fe$_{0.05}$O$_3$ at a few typical temperatures. The sharp mode at ~ 308 cm$^{-1}$ is a Raman phonon of the tetragonal BaTiO$_3$.

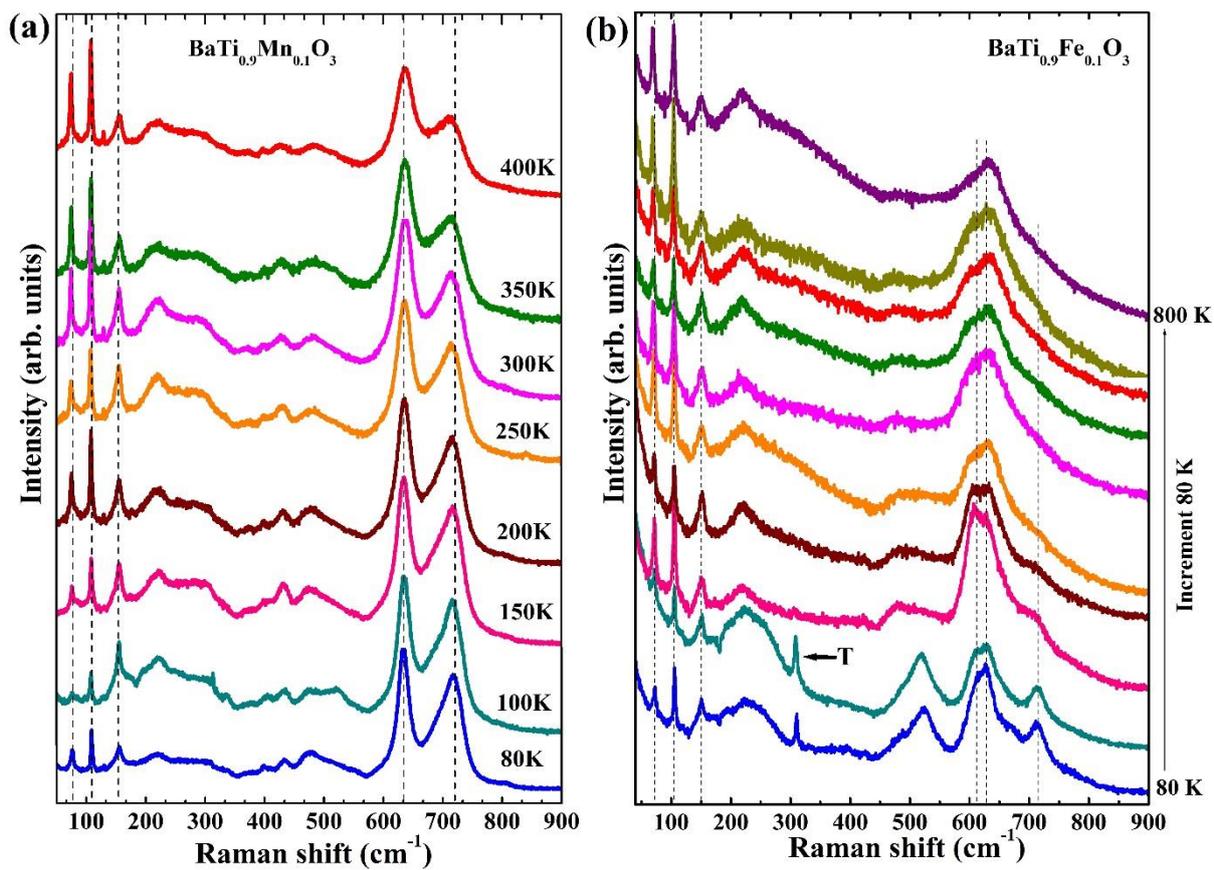

Figure S9. Raman spectra of BaTi$_{0.9}$Mn$_{0.1}$O$_3$ and BaTi$_{0.9}$Fe$_{0.1}$O$_3$ at a few typical temperatures. The sharp mode at ~ 308 cm$^{-1}$ is a Raman phonon of the tetragonal BaTiO$_3$.

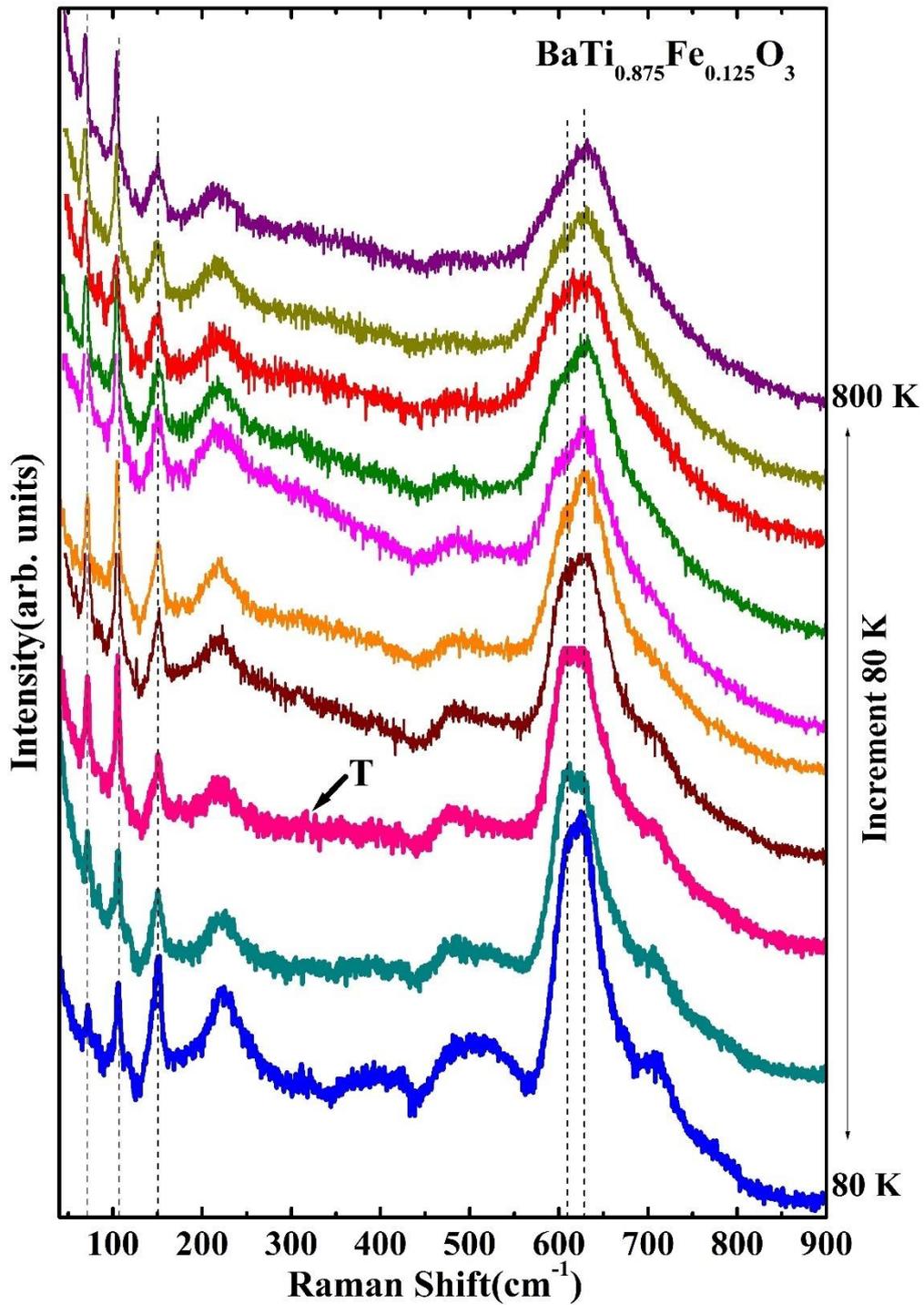

Figure S10. Raman spectra of BaTi$_{0.875}$Fe$_{0.125}$O$_3$ at a few typical temperatures. The mode at ~ 308 cm$^{-1}$ is a Raman phonon of the tetragonal BaTiO$_3$.

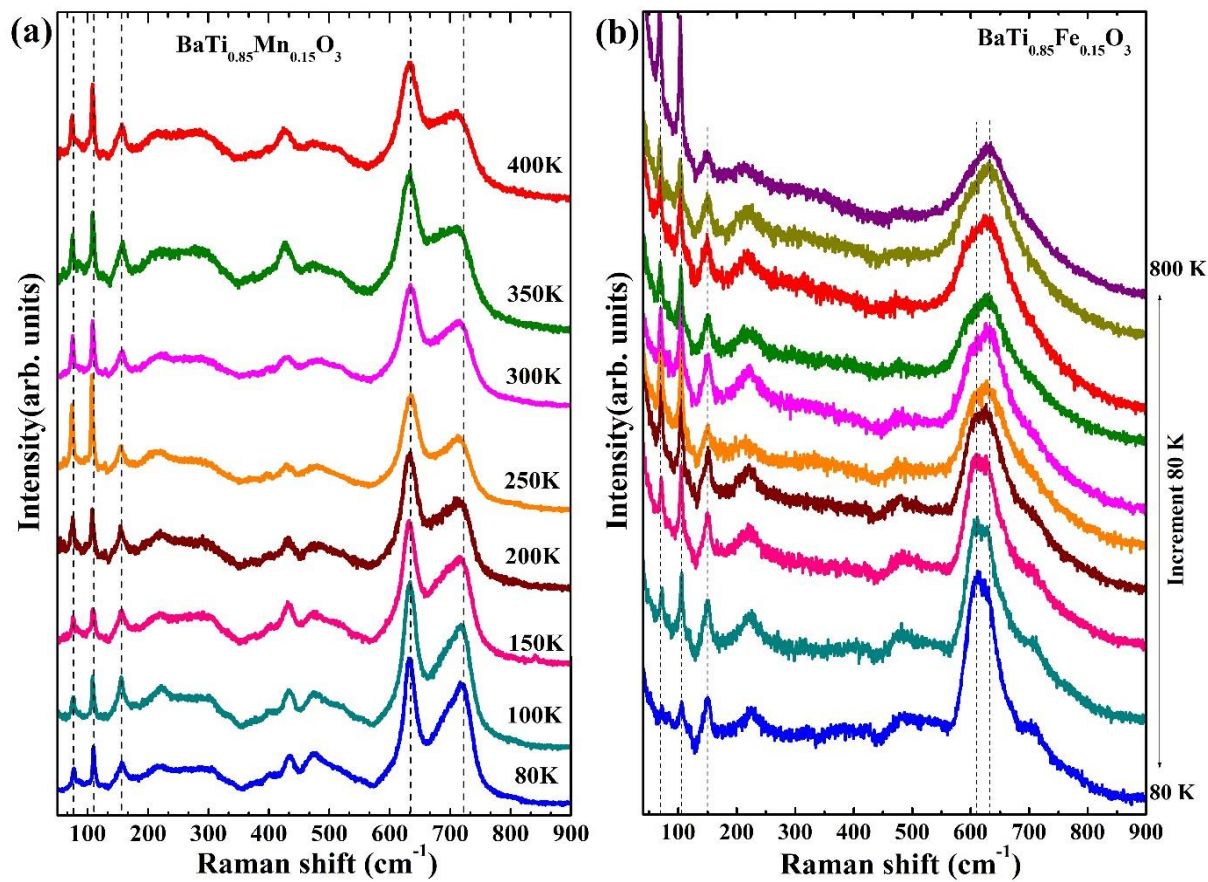

Figure S11. Raman spectra of $BaTi_{0.85}Mn_{0.15}O_3$ and $BaTi_{0.85}Fe_{0.15}O_3$ at a few typical temperatures.

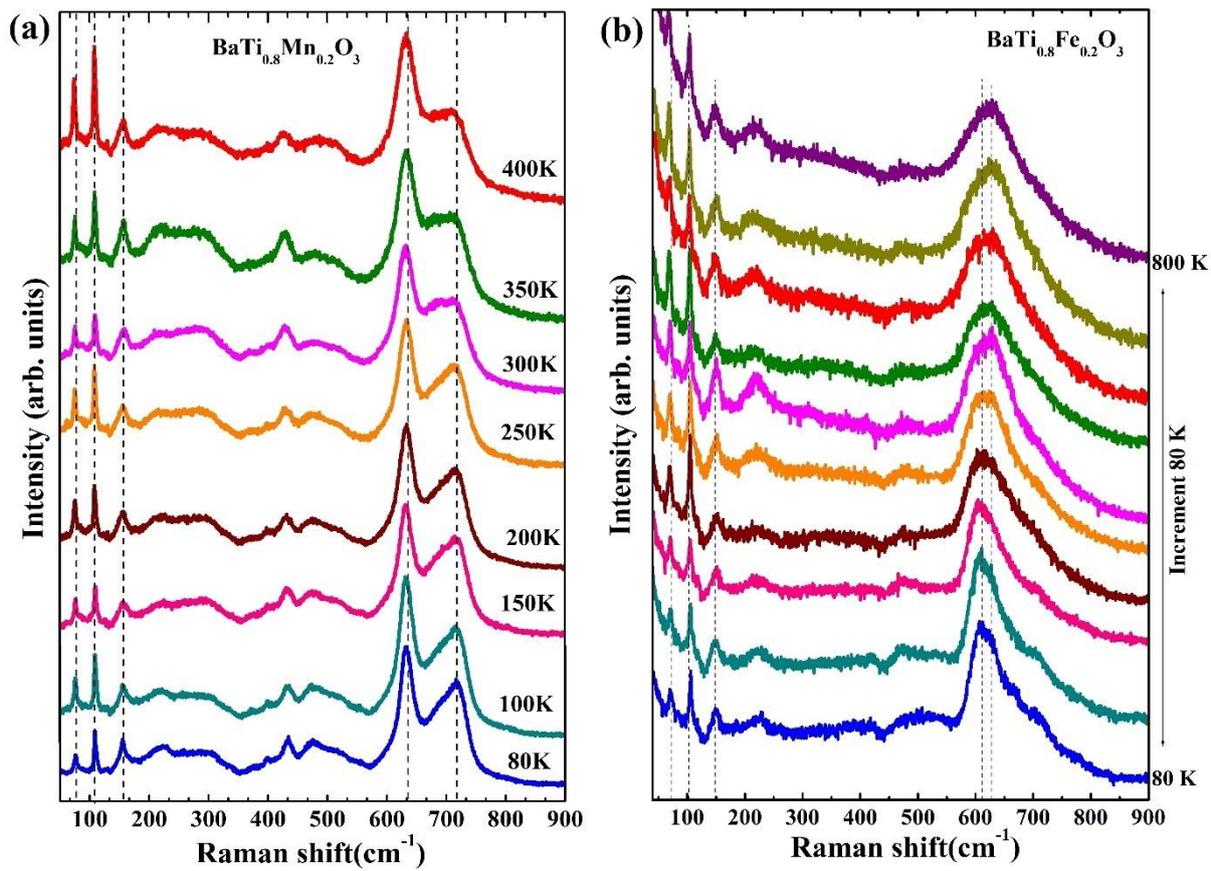

Figure S12. Raman spectra of BaTi$_{0.8}$Mn$_{0.2}$O$_3$ and BaTi$_{0.8}$Fe$_{0.2}$O$_3$ at a few typical temperatures.

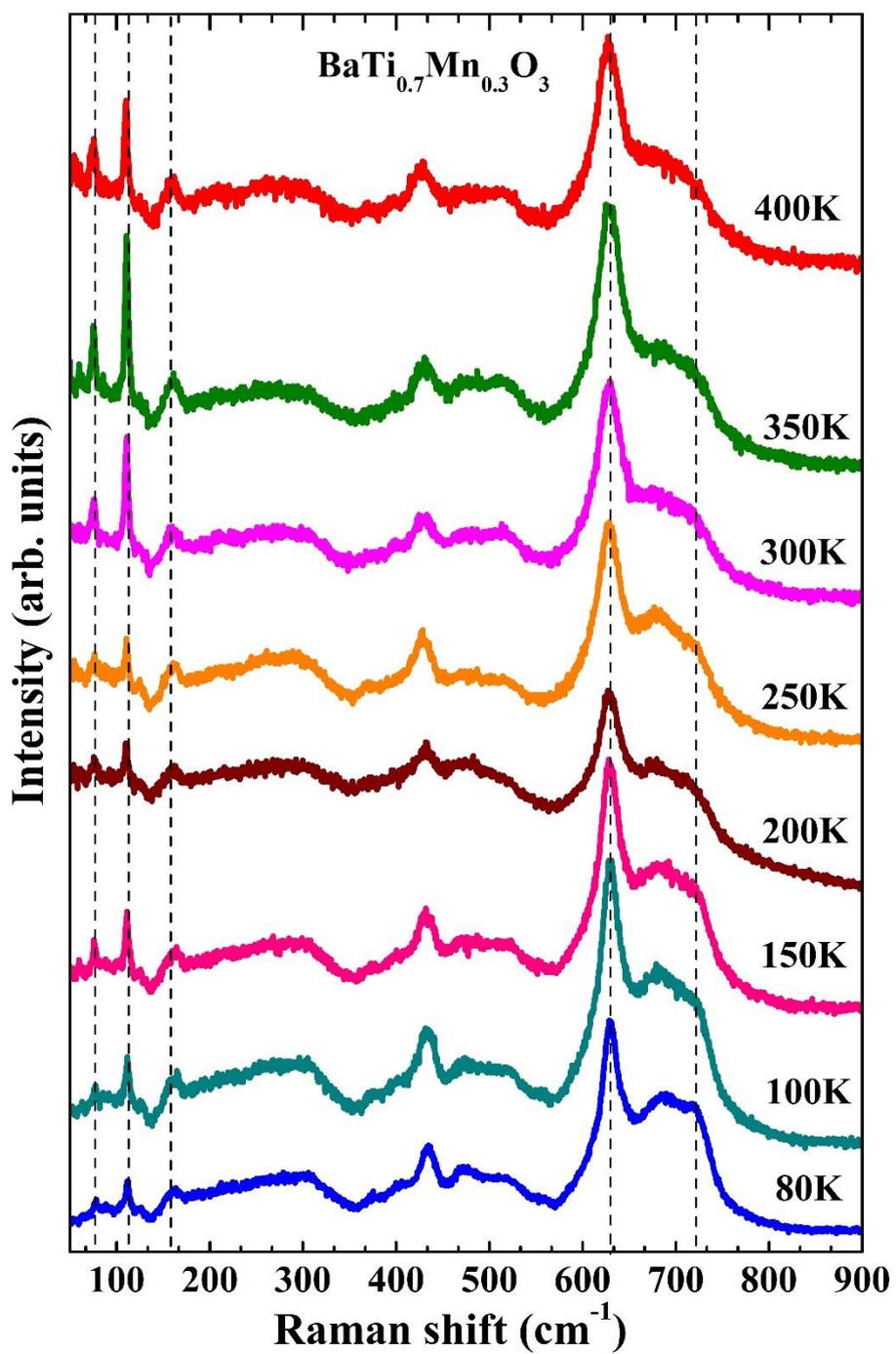

Figure S13. Raman spectra of BaTi$_{0.7}$Mn$_{0.3}$O$_3$ at a few typical temperatures.

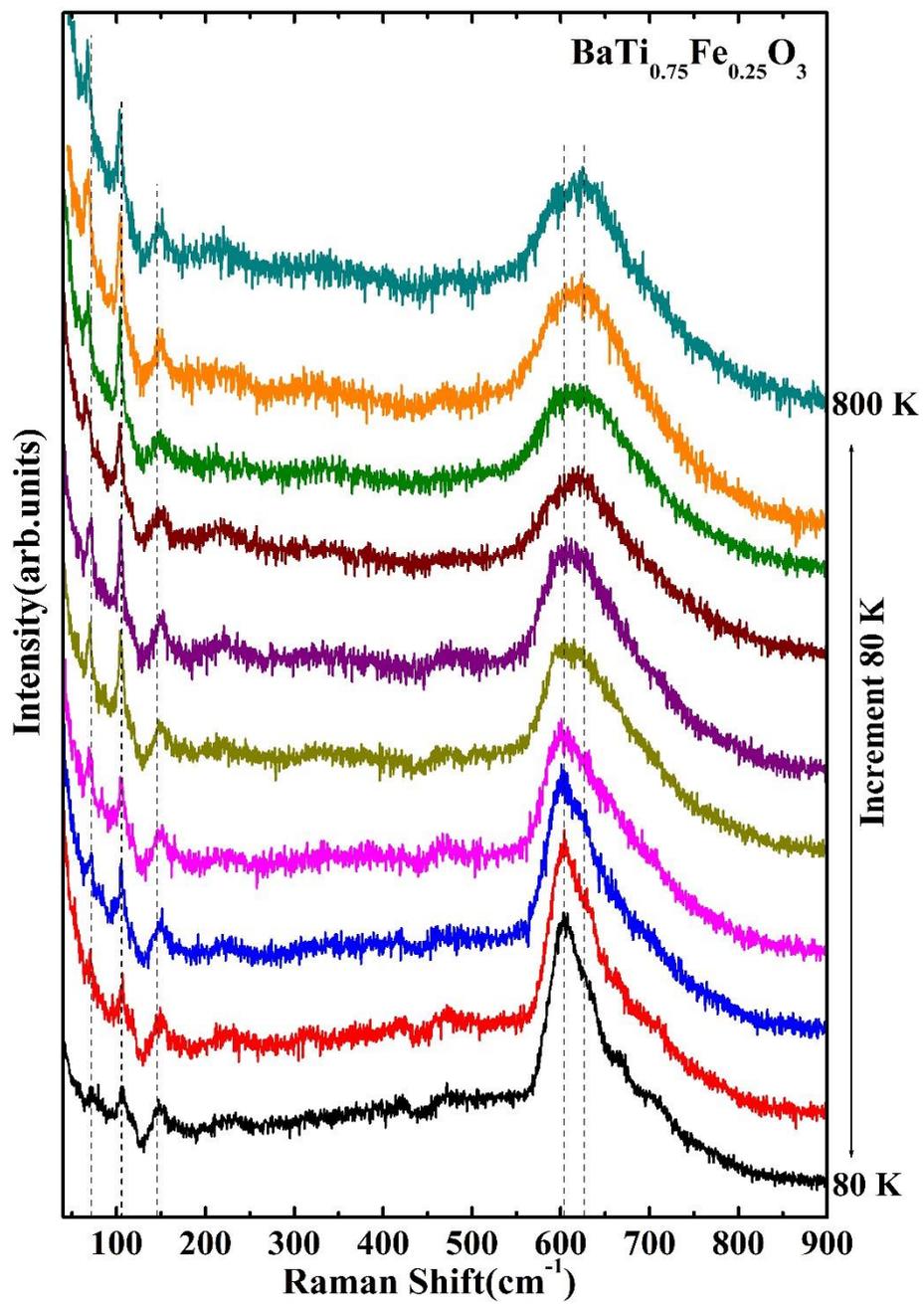

Figure S14. Raman spectra of BaTi$_{0.75}$Fe$_{0.25}$O$_3$ at a few typical temperatures.

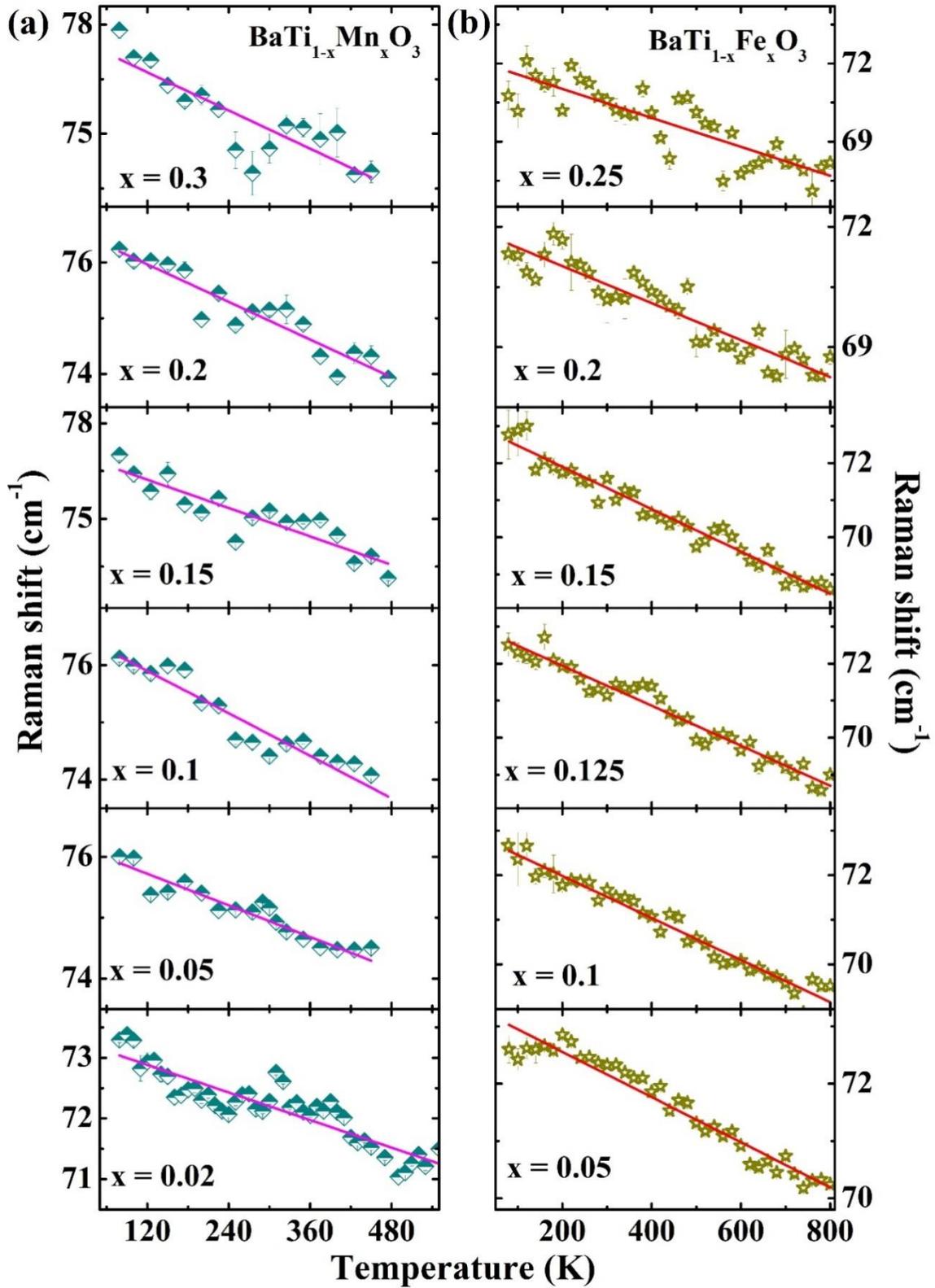

Figure S15. $\omega$ vs T for $E_{2g}$ mode at 71 cm$^{-1}$. Solid lines represent fitting with Eq. 2 explained in the main text.

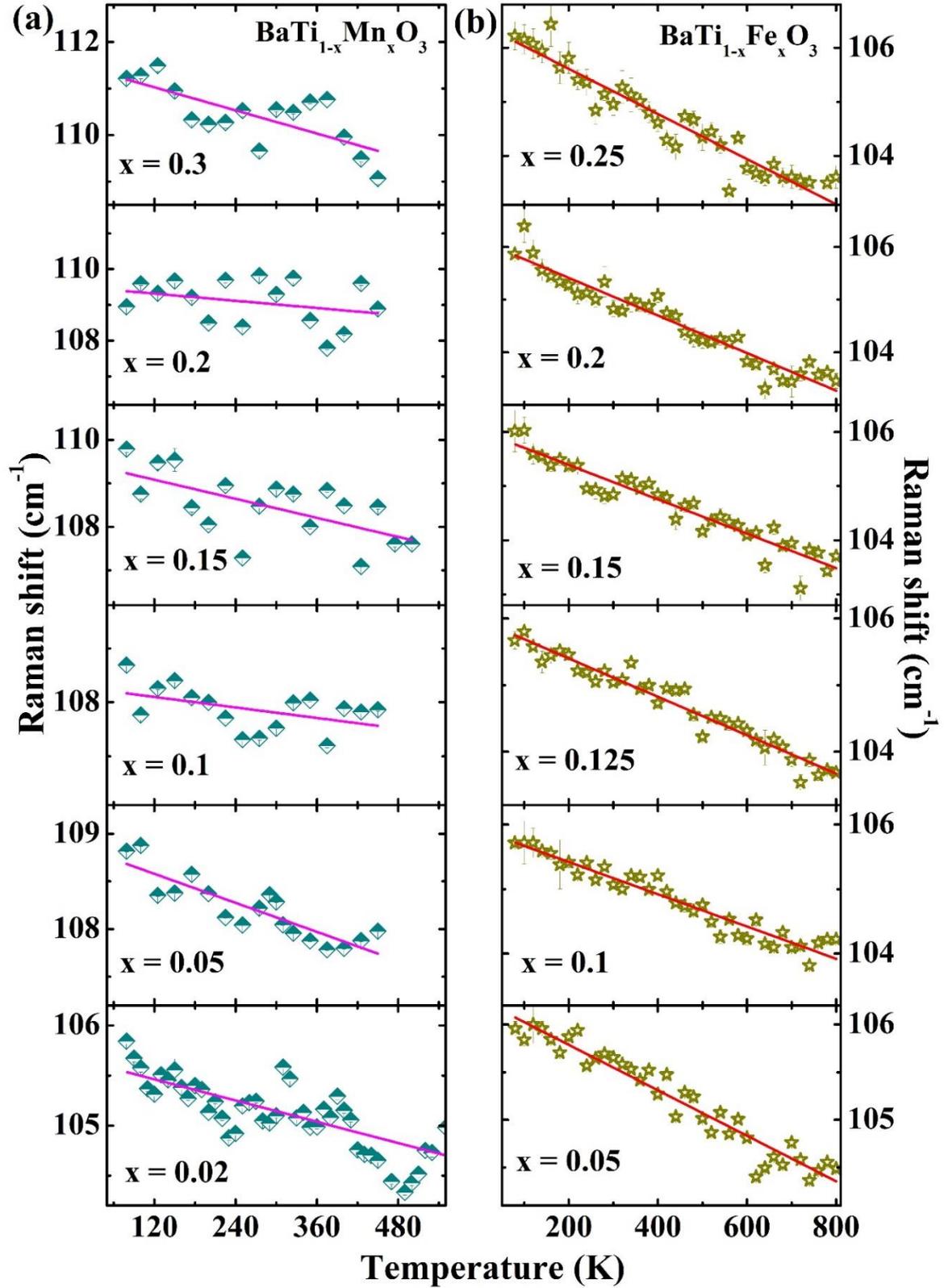

Figure S16. $\omega$ vs T for $A_{1g}$ mode at 104 cm$^{-1}$. Solid lines represent fitting with Eq. 2 explained in the main text.

## S6. Thermal expansion: Temperature-dependent x-ray diffraction

We have done temperature-dependent x-ray diffraction measurements for $BaTi_{0.98}Mn_{0.02}O_3$, $BaTi_{0.7}Mn_{0.3}O_3$, $BaTi_{0.875}Fe_{0.125}O_3$, and $BaTi_{0.85}Fe_{0.15}O_3$ in the range of 90 - 410 K. The refined lattice parameters correspond to 6H phase as a function of temperature are shown in Figure S17 and S18.

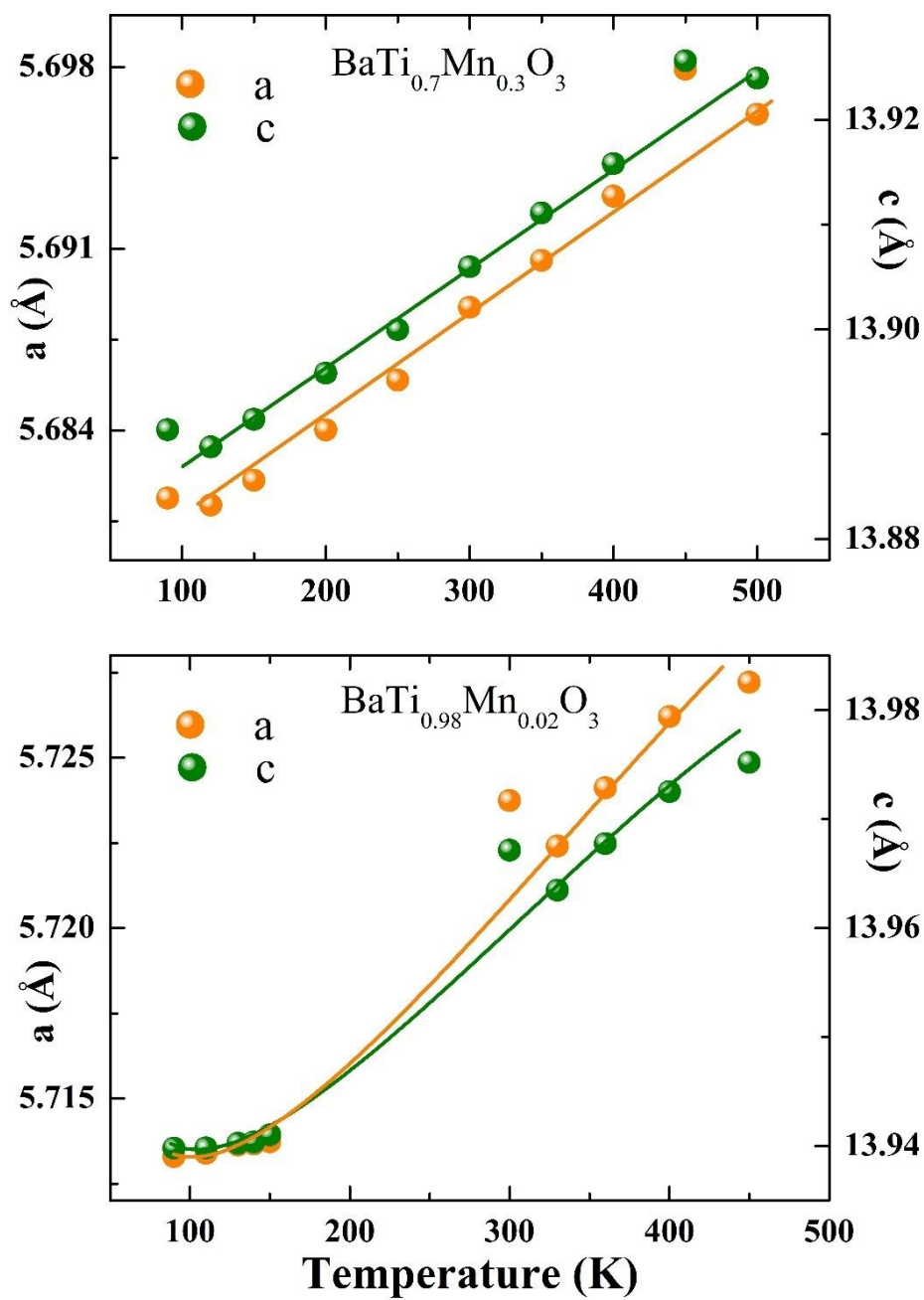

Figure S17. Temperature-dependent lattice parameters of $BaTi_{0.98}Mn_{0.02}O_3$ and $BaTi_{0.7}Mn_{0.3}O_3$.

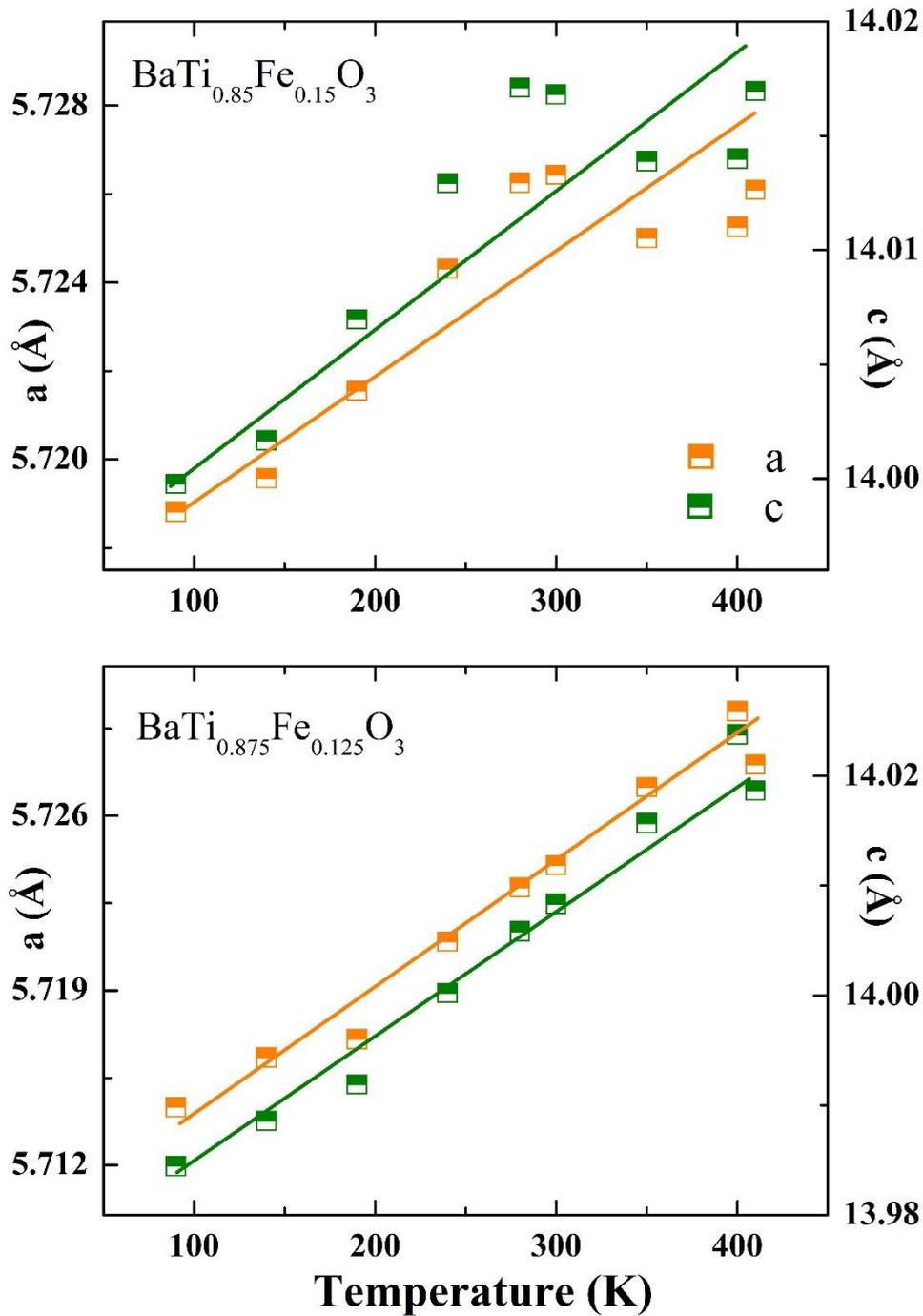

Figure S18. Temperature-dependent lattice parameters of BaTi$_{0.875}$Fe$_{0.125}$O$_3$ and BaTi$_{0.85}$Fe$_{0.15}$O$_3$. Solid lines are guide to eye.

### S7. Magnetism

As also discussed in the main text (Figure 6(a)), BaTi$_{1-x}$Mn$_x$O$_3$: x ≥0.15 exhibit a magnetic transition around T* ~ 43 K. As temperature is lowered, ZFC and FC curves in M(T) show significant bifurcation below T* ~ 43 K exhibiting a peak-like feature in ZFC curve. This feature indicates that the magnetic phase below T* can be ferrimagnetic or canted antiferromagnetic or a glassy type (spin glass or cluster glass) phase. However, based on the previous reports [6-8], we attribute this to ferrimagnetic or canted antiferromagnetic transition.

Figure S19 displays inverse magnetic susceptibility as a function of temperature. The Curie-Weiss constant ($\theta_C$) extracted from the fitting is plotted against doping (Fe) in Figure S20. The in-plane lattice parameter, magnetic moment, coercive field, and Curie-Weiss constant show a similar trend as a function of doping (Fe) in BaTi$_{1-x}$Fe$_x$O$_3$ system representing the strong correlation among structural, magnetic, ferroelectric properties, and lattice strain. Bond angles M-O-M are plotted as a function of doping in Figure S22. While M1-O2-M2 (~ 177º) decreases with doping, the angle M2-O2-M2 increases which imply that the antiferromagnetic interactions decrease and ferromagnetic interactions increase with doping (Mn/Fe) in both systems.

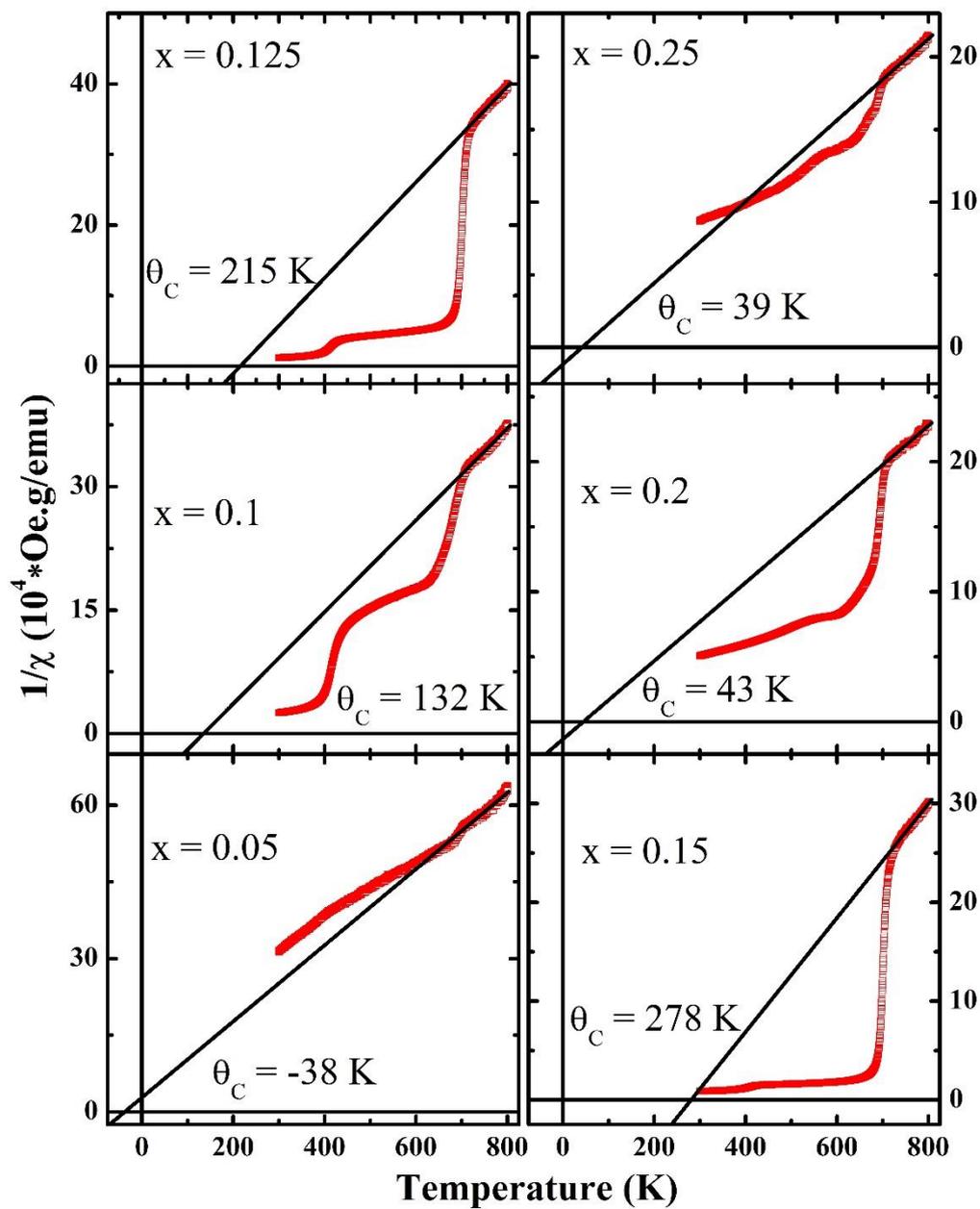

Figure S19. Inverse magnetic susceptibility as a function of temperature for BaTi$_{1-x}$Fe$_x$O$_3$: x = 0.05 to 0.25.

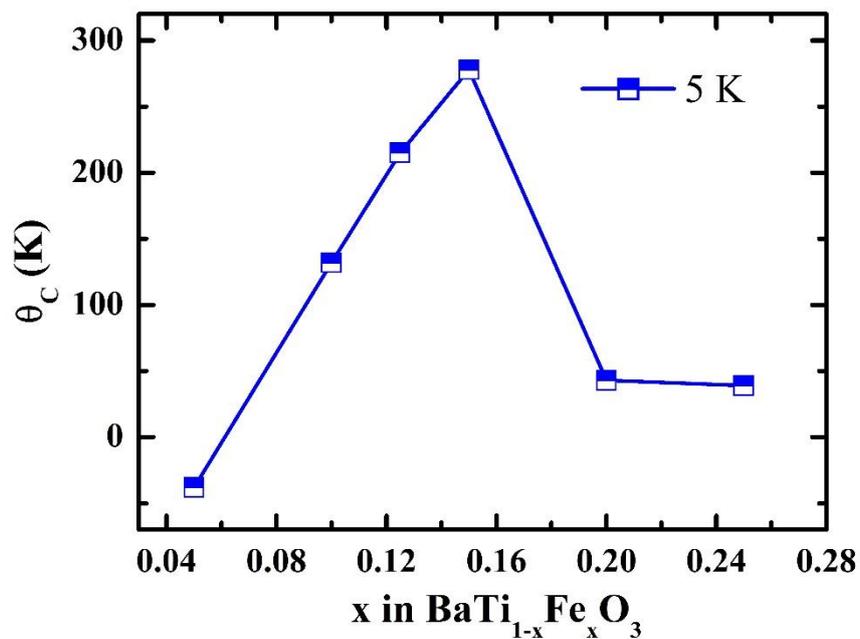

Figure S20. Curie-Weiss temperature ($\theta_C$) as a function of doping in $BaTi_{1-x}Fe_xO_3$.

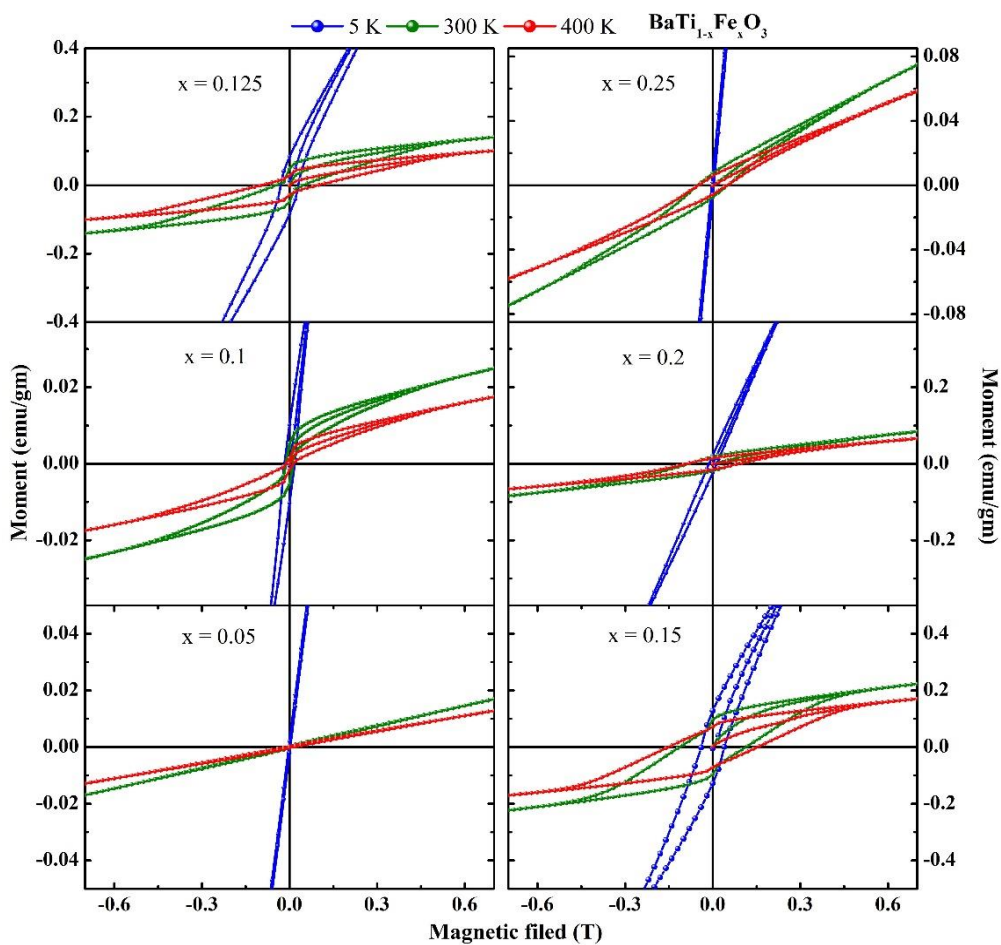

Figure S21. Magnetization as a function of magnetic field at a few temperatures for $BaTi_{1-x}Fe_xO_3$: x = 0.05 to 0.25.

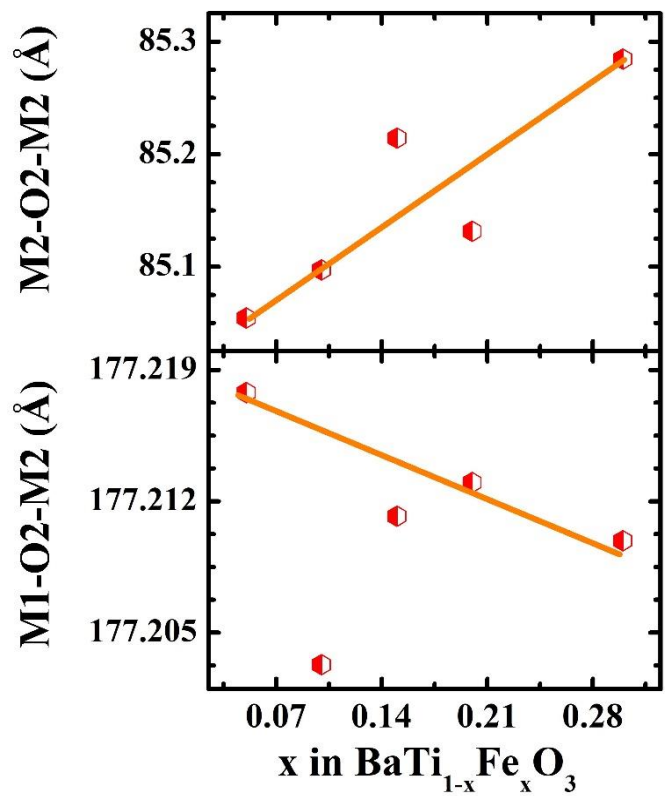
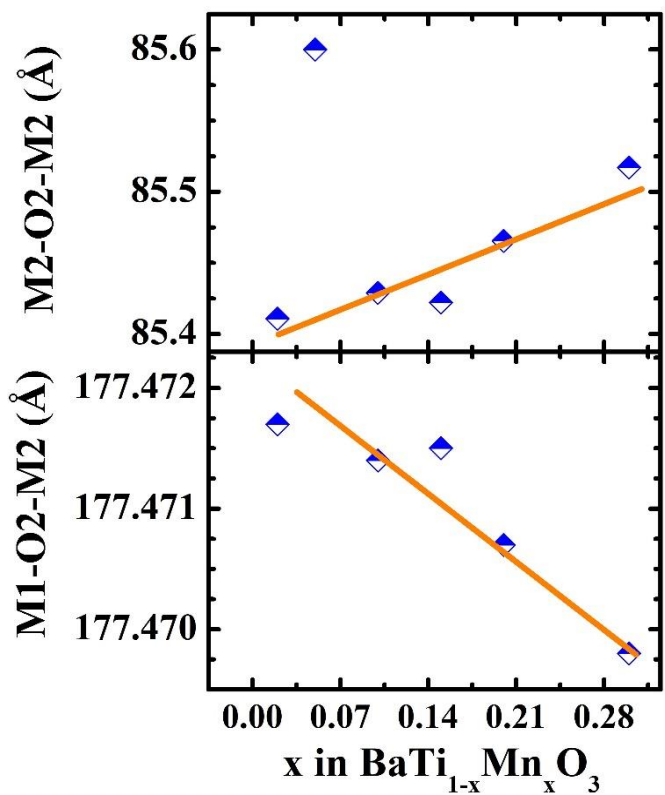

Figure S22. Bond angles as a function of doping (Mn/Fe) in BaTi$_{1-x}$Mn$_x$O$_3$ and BaTi$_{1-x}$Fe$_x$O$_3$. Solid lines are guide to eye.

## S8. Phonon anharmonicity

The phonon anomaly is quantified using the relation $\frac{\Delta\omega}{\omega}(\%) = 100 * \frac{\omega_{400K}-\omega_{80K}}{\omega_{80K}}$ and plotted in Figure S23 as a function of doping (Mn/Fe) for $E_{1g}$ (at 152 cm$^{-1}$) and $A_{1g}$ (at 636 cm$^{-1}$) modes. The phonon anomaly decreases with increasing Mn/Fe concentration at Ti-site due to the suppression of anharmonic phonon-phonon interactions. The difference in the anomaly between the two systems can be related to the difference in their bond lengths and lattice parameters. The Ba-O bond length decreases (increases) with Mn (Fe) concentration which in turn increases (decreases) the quasi-harmonic contribution (associated with the change in volume) to the phonon frequency in BaTi$_{1-x}$Mn$_x$O$_3$ (BaTi$_{1-x}$Fe$_x$O$_3$) system. The quasi-harmonic contribution is insufficient to explain the observed decreasing trend of an anomaly with Mn/Fe concentration in both systems. Hence, we conclude that the partial substitution of Mn/Fe at Ti site suppresses the anharmonic phonon-phonon interactions. The $A_{1g}$ [635 cm$^{-1}$] mode in BaTi$_{1-x}$Mn$_x$O$_3$ is the convolution of two phonons: one arises from non-magnetic Ti ion (Ti-O) vibrations and the other is related to the displacement of magnetic (Mn) ions. As a result, the combined effect is present in the temperature-dependence of it. The $A_{1g}$ phonon of frequency ~ 637 cm$^{-1}$ in BaTi$_{1-x}$Fe$_x$O$_3$ is solely because of Ti(Ti-O) vibrations showing anomalous behaviour. Another $A_{1g}$ phonon of frequency ~ 600 cm$^{-1}$ appears as a new mode which is associated with Fe (Fe-O) vibrations alone, following the anharmonic model (Eq. 2). This observation leads to the conclusion that the anomalous temperature dependence arises from non-magnetic ions (Ti) in both systems. The temperature-dependence of Ti and Mn vibrations is opposite to each other, where the former leads frequency to increase and the latter causes it to decrease with temperature resulting the $A_{1g}$ [635 cm$^{-1}$] mode in BaTi$_{1-x}$Mn$_x$O$_3$ to have less anomaly as compared to BaTi$_{1-x}$Fe$_x$O$_3$ system. The anomaly decreases with (Mn/Fe) doping due to suppression of phonon-phonon interactions (quantified using the relation $\frac{\Delta\omega}{\omega} = \frac{\omega_{400K}-\omega_{80K}}{\omega_{80K}}$ shown in Figure S23). It should be noted here that these phonons do not get renormalized around ferromagnetic transition (T$_{C1}$) in BaTi$_{1-x}$Fe$_x$O$_3$ and hence, the presence of spin-phonon coupling is ruled out. Thus, we attribute the origin of phonon anomalies to strong anharmonic phonon-phonon interactions which get suppressed by doping (Mn/Fe) at Ti-site.

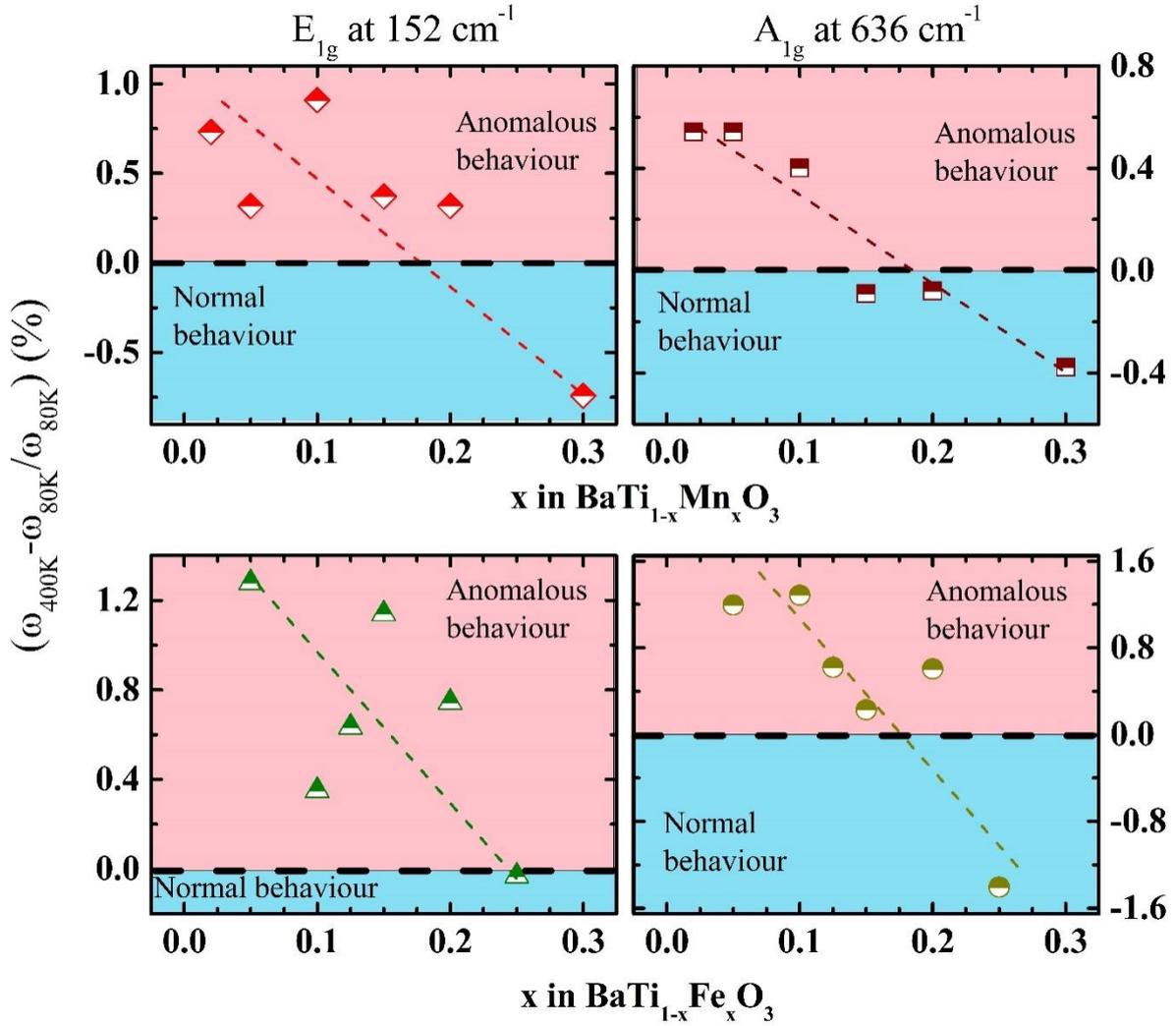

Figure S23. Plot of $\frac{d\omega}{\omega} = \frac{\omega_{400K}-\omega_{80K}}{\omega_{80K}}$ (%) as a function of doping (Mn/Fe) of $E_{1g}$ mode at 152 cm$^{-1}$ and $A_{1g}$ mode at 636 cm$^{-1}$. The values of $\frac{d\omega}{\omega} > 0$ signify an anomalous behaviour of the phonon while the $\frac{d\omega}{\omega} < 0$ corresponds to normal behaviour.

References

1. Y. I. Yuzyuk, "Raman scattering spectra of ceramics, films, and superlattices of ferroelectric perovskites: A review", Phys. Solid State 54, (2012) 1026-1059. https://doi.org/10.1134/S1063783412050502
2. G. Pezzottia, "Raman spectroscopy of piezoelectrics", J. Appl. Phys. **113**, (2013) 211301. http://dx.doi.org/10.1063/1.4803740
3. L. Rimai, J. L. Parsons, and J. T. Hickmott, "Raman Spectrum of Long-Wavelength Phonons in Tetragonal Barium Titanate", Phys. Rev. **168**, (1968) 623. https://doi.org/10.1103/PhysRev.168.623